\documentclass[3p,times]{elsarticle}
\pdfoutput=1

\usepackage[version=4]{mhchem} 
\usepackage{graphicx}
\usepackage{subcaption}
\usepackage{multirow}
\usepackage{siunitx}
\usepackage{textcomp}
\usepackage{color}
\usepackage{tikz}
\usepackage{wrapfig}
\usepackage[strings]{underscore}
\makeatletter
\def\ps@pprintTitle{%
 \let\@oddhead\@empty
 \let\@evenhead\@empty
 \def\@oddfoot{\centerline{\thepage}}%
 \let\@evenfoot\@oddfoot}
\makeatother

\begin{document}

\begin{frontmatter}


\title{A  First Principles Investigation of Native Interstitial Diffusion in \ce{Cr2O3}}

    \author[1]{Bharat Medasani}
    \author[1]{Maria L. Sushko}
    \author[1]{Kevin M. Rosso}
    \author[2]{Daniel K. Schreiber}
    \author[2]{Stephen M. Bruemmer}
    \cortext[author]{Corresponding author: B. Medasani (bharat.medasani@pnnl.gov, mbkumar@gmail.com). }

    \address[1]{Physical and Computational Sciences Directorate, Pacific Northwest National Laboratory, Richland, WA 99354, USA }
    \address[2]{Energy and Environment Directorate, Pacific Northwest National Laboratory, Richland, WA 99354, USA }


\begin{abstract}
  First principles density functional theory (DFT) investigation of native interstitials and the associated self-diffusion mechanisms in $\alpha$-\ce{Cr2O3} reveals that interstitials are more mobile than vacancies of corresponding species. Cr interstitials 
occupy the unoccupied Cr sublattice sites that are octahedrally coordinated by 6 O atoms,  and 
O interstitials form a dumbbell configuration orientated along the [221] direction (diagonal) of the 
corundum lattice.  Calculations 
predict that neutral O interstitials 
are predominant in O-rich conditions and  Cr interstitials in +2 and +1 charge states are
the dominant interstitial defects in Cr-rich conditions. Similar to that of the vacancies, the charge 
transition levels of both O and Cr interstitials are located deep within the bandgap.
 Transport calculations reveal a rich variety of interstitial diffusion mechanisms that are  
species, charge, and orientation dependent.  
Cr interstitials diffuse preferably along the diagonal of corundum lattice 
in a two step process via an intermediate defect  complex comprising  a Cr interstitial and 
an adjacent Cr  Frenkel  defect in the neighboring Cr bilayer. This mechanism is 
similar to that of the vacancy 
mediated Cr  diffusion along the c-axis with intermediate Cr vacancy and 
Cr Frenkel defect combination.  In contrast, O interstitials diffuse via bond 
switching mechanism.  O interstitials in -1 and -2 charge states have very high 
mobility compared to neutral O interstitials.
\end{abstract}
\end{frontmatter}


\section{Introduction}
The global cost of corrosion is estimated to be US \$2.5 trillion annually~\cite{naceimpact}. To prevent corrosion, many
Ni and Fe-based alloys utilize Cr as an alloying element
due to its ability to form \ce{Cr2O3} as a protective oxide 
coating~\cite{infacon1992,rooyen1992,asm,was2005,zinkle2009}. These alloys are often 
used in extreme operating conditions such as in nuclear 
reactors, where the irradiating conditions can create
defects within the protective oxide film, or in turbine blades with high operating temperatures. 
The effectiveness of these alloys under such extreme conditions 
has  been the subject of many experimental studies~\cite{schreiber2013,schreiber2014}.
Understanding the mechanism and the limits of corrosion 
protection offered by \ce{Cr2O3} requires detailed atomistic 
models of defect formation and transport in this oxide. 
Towards this objective, many experimental  studies of defects, 
oxidation mechanisms,  and diffusion in \ce{Cr2O3} were carried out\cite{kofstad1982,sabioni1992a,sabioni1992b,sabioni1992c,hoshino1983,latanision1967,Schmucker2016}. 
Some of these studies indicated that Cr interstitials 
could be the  dominant defects in \ce{Cr2O3} under 
reducing  conditions~\cite{latanision1967,Schmucker2016}.
Classical Monte Carlo simulations also predicted that Cr interstitials had lower
barrier energies and, therefore, higher mobility than vacancies~\cite{cao2017}.
These studies suggest that interstitials could be controlling 
the effectiveness of \ce{Cr2O3} as a protective layer in 
pressurized water reactors (PWR), where reducing conditions 
prevail due to the use  of hydrogenated water as a coolant.

Advances in experimental, theoretical and 
computational  methods led to a renewed  focus on defects 
and their diffusion mechanisms in \ce{Cr2O3} in recent 
years\cite{Lebreau2014,medasani2017,gray2016,cao2017,Schmucker2016,Vaari2015,laturomain2017}.  
Within the last decade,  first principles density functional 
theory (DFT)~\cite{hohenberg, kohn,dftprimer} has emerged as a reliable 
tool to model  point defects in bandgap materials in part due to 
the development  of \textit{a posteriori} correction 
techniques~\cite{freysoldt2014}. However,
DFT simulations of defects and their diffusion in \ce{Cr2O3} 
were focused  exclusively on 
vacancies~\cite{Lebreau2014,medasani2017,gray2016}, while  
diffusion of interstitials defects has not been studied. To obtain a 
comprehensive atomistic picture of native defects and their diffusion in \ce{Cr2O3},
here,  we investigated Cr and O interstitials in 
\ce{Cr2O3}  in the charge states ranging between [0, 3] and 
[-2, 0], respectively, 
using density functional theory and identified their diffusion mechanisms and pathways.
 
\section{Methods\label{sec:methods}}
\subsection{Interstitial Site Selection}
\begin{wraptable}{l}{0.6\textwidth}
    \centering
\renewcommand{\arraystretch}{1.4}
\caption{Correspondence between the Kroger-Vink (K-V)
representation and the representation used in this 
work for charged interstitials.}
\begin{tabular}{lccc}
\hline
Defect & Charge & K-V Notation & Notation in this study \\
\hline
\multirow{4}{*}{Cr interstitial} & 0 & $Cr_i^X$ & $Cr_i^0$ \\
  & 1 & $Cr_i^\bullet$ & $Cr_i^1$ \\
  & 2 & $Cr_i^{\bullet\bullet}$ & $Cr_i^{2}$ \\
  & 3 & $Cr_i^{\bullet\bullet\bullet}$ & $Cr_i^{3}$ \\ \hline
\multirow{3}{*}{O interstitial}& 0 & $V_O^X$ & $V_O^0$ \\
  & -1 & $O_i^{'}$ & $O_i^{-1}$ \\
  & -2 & $O_i^{''} $ & $O_i^{-2}$ \\
\hline
\end{tabular}
\label{tab:internotation}
\end{wraptable}
This work is a continuation from our previous study of 
vacancies and vacancy  mediated self-diffusion in \ce{Cr2O3}~\cite{medasani2017}. Therefore, the methods and associated  parameters used
in this study are the same as 
those in our study on the vacancies in \ce{Cr2O3}. The notation used in 
this work  for interstitials with different charge states and the corresponding 
Kroger-Vink notation are provided in Table~\ref{tab:internotation}.
We used PyCDT~\cite{pycdt} and pymatgen~\cite{ong2012b} to identify potential interstitial sites in \ce{Cr2O3}. 
The interstitial finding procedure~\cite{zimmermann2016} implemented in pymatgen systematically performs a grid based search for potential interstitial sites by evaluating the coordination pattern at each grid point. Grid points that exhibit the
coordination patterns resembling basic structural motifs (e.g., tetrahedral and octahedral environments) with high symmetry are identified by this algorithm as potential interstitial sites.  

\begin{figure}
\includegraphics[width=1\linewidth]{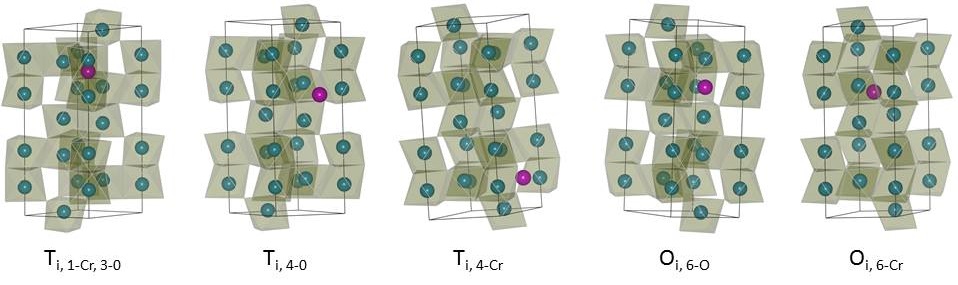}
\caption{Potential interstitial sites (magenta spheres) for \ce{Cr2O3} obtained from PyCDT~\cite{pycdt}. $T_i$ and $O_i$ designations represent tetrahedral and octahedral sites respectively. The subscript identifies the type and number of coordinating elements. Teal spheres represent Cr atoms and the O atoms are at the vertices of the octahedra surrounding the Cr atoms.}
\label{fig:interstitialsites}
\end{figure}

\subsection{Density Functional Calculations\label{subsec:methods-dft}}
Interstitials were modeled with periodic supercell formalism. For charged 
interstitials, neutralizing background charge was automatically added by the 
DFT software, making the supercells neutral. 
Geometry relaxation, barrier energy, and phonon calculations for this study were performed within the DFT framework implemented in the
Vienna \textit{ab initio} simulation package (VASP) \cite{kresse93,kresse94,kresse96}. 
We included Hubbard on-site Coulombic correction (U) using GGA+U
method\cite{dftuliech,Rohrbach2004}. The basis sets of Cr and O consist of 12 
and 6 valence electrons in the configurations of [He]2s$^2$2p$^4$ and
[Ne$\,$3s$^2$]3p$^6$4s$^2$3d$^4$  respectively. 
Further details of GGA+U simulations are given in Supplementary Information (SI).
The DFT settings used in this study were unchanged from those used in our 
previous work  on vacancies in \ce{Cr2O3}~\cite{medasani2017}, and the bulk properties 
of \ce{Cr2O3} obtained with 
these settings can be found in our previous publication.~\cite{medasani2017}

We used the thermodynamic formalism proposed by Zhang and Northrup~\cite{zhang1991} to evaluate the 
formation energies of the interstitials. The formation energy of an interstitial of 
species $X$ with charge $q$, $E_f(X^q_i)$, was calculated as 
\begin{equation}
E_f(X_i^q) = E_{tot}(X_i^q) - E_{tot}[bulk] - \mu_X + qE_F + E_{corr},
\label{eq:interformen}
\end{equation}
where $E_{tot}(X_i^q)$ is the total energy of the defect supercell, $E_{tot}[bulk]$ 
is the total energy of the bulk supercell, $\mu_X$ is the chemical potential 
contribution resulting from the removal of the atom/ion of specie X, $E_F$ is the Fermi 
level and $E_{corr}$ is the correction term, which encompasses the 
correction to finite-size electrostatic interactions and bandgap underestimation in DFT~\cite{medasani2017}. The barrier energies were computed using 
CI-NEB method~\cite{cineb}. 

The stability of predicted interstitial sites 
and transition states  was evaluated using phonon density 
of states (PDOS) calculations. The PDOS of stable
interstitial sites should 
exhibit no imaginary frequencies, while the transition states, 
being saddle points, should have one imaginary frequency in their 
PDOS. Additional details about the phonon calculations are provided in the SI.
\begin{figure}
  \begin{subfigure}{.5\textwidth}
    \centering
    \includegraphics[width=1\linewidth]{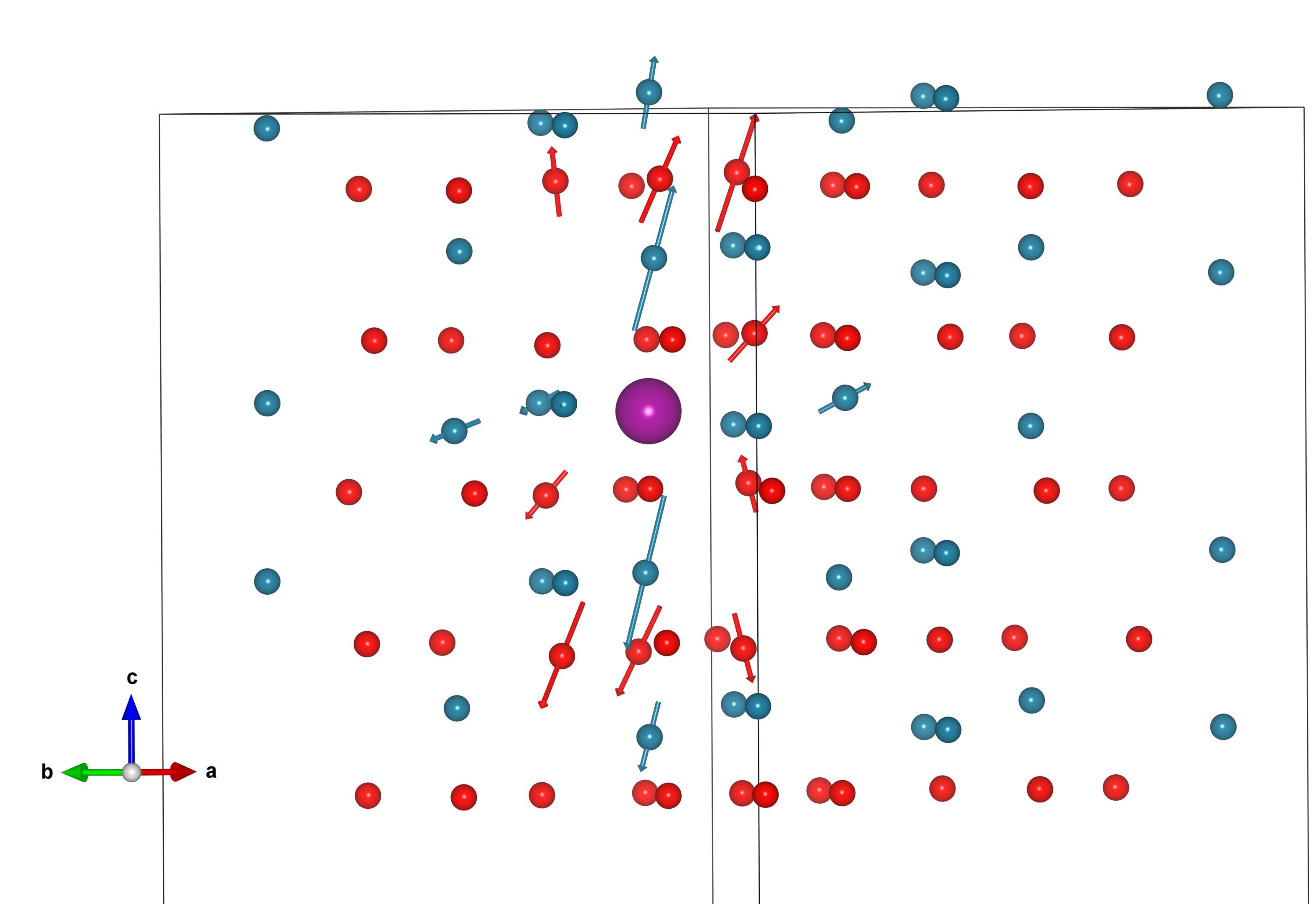}
    \caption{${Cr}_i^0$}
    \label{fig:Crinterdispvectq0}
  \end{subfigure}%
  \begin{subfigure}{.5\textwidth}
    \centering
    \includegraphics[width=1\linewidth]{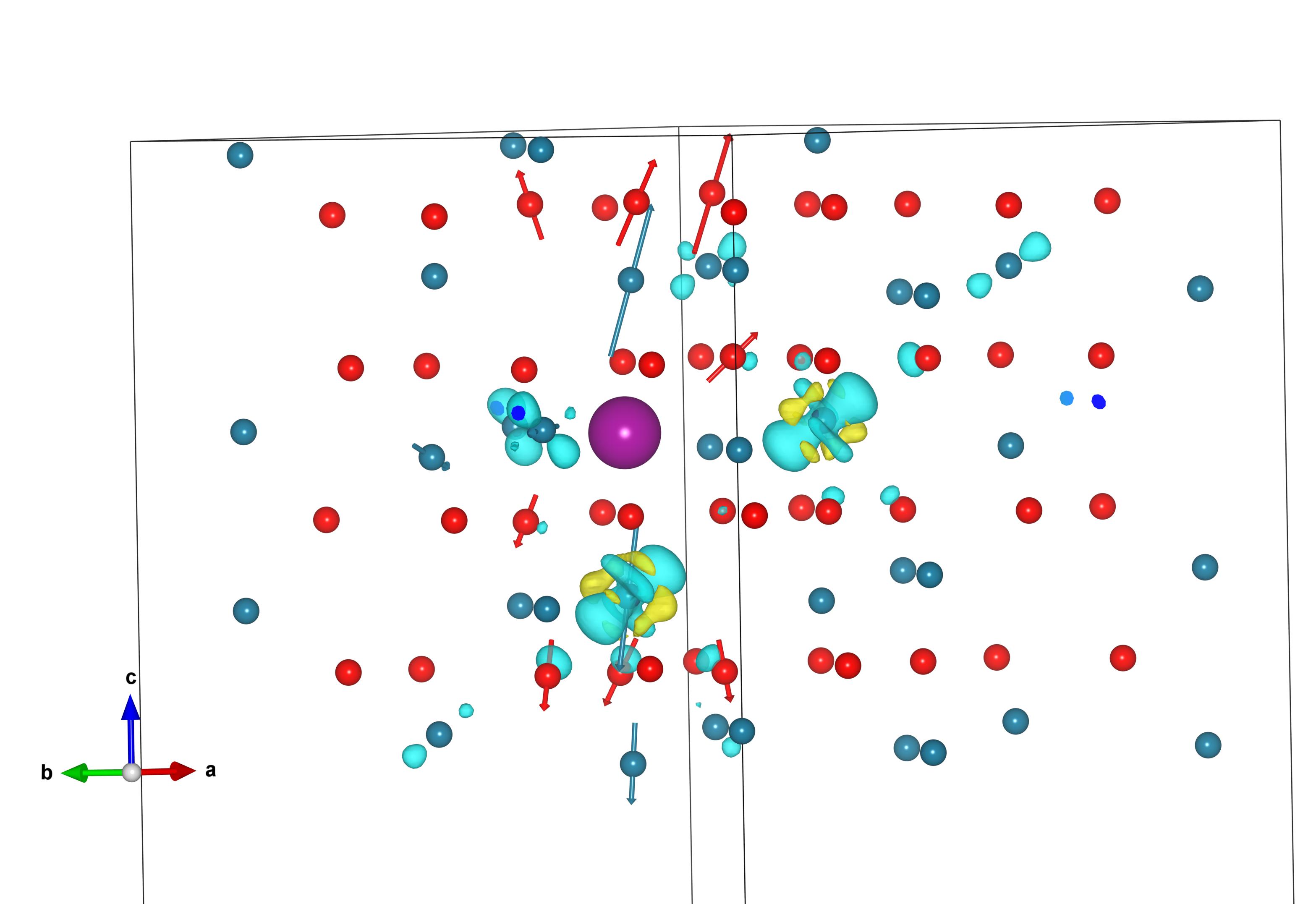}
    \caption{${Cr}_i^{+1}$}
    \label{fig:Crinterdispvectq1}
  \end{subfigure}
  \begin{subfigure}{.5\textwidth}
    \centering
    \includegraphics[width=1\linewidth]{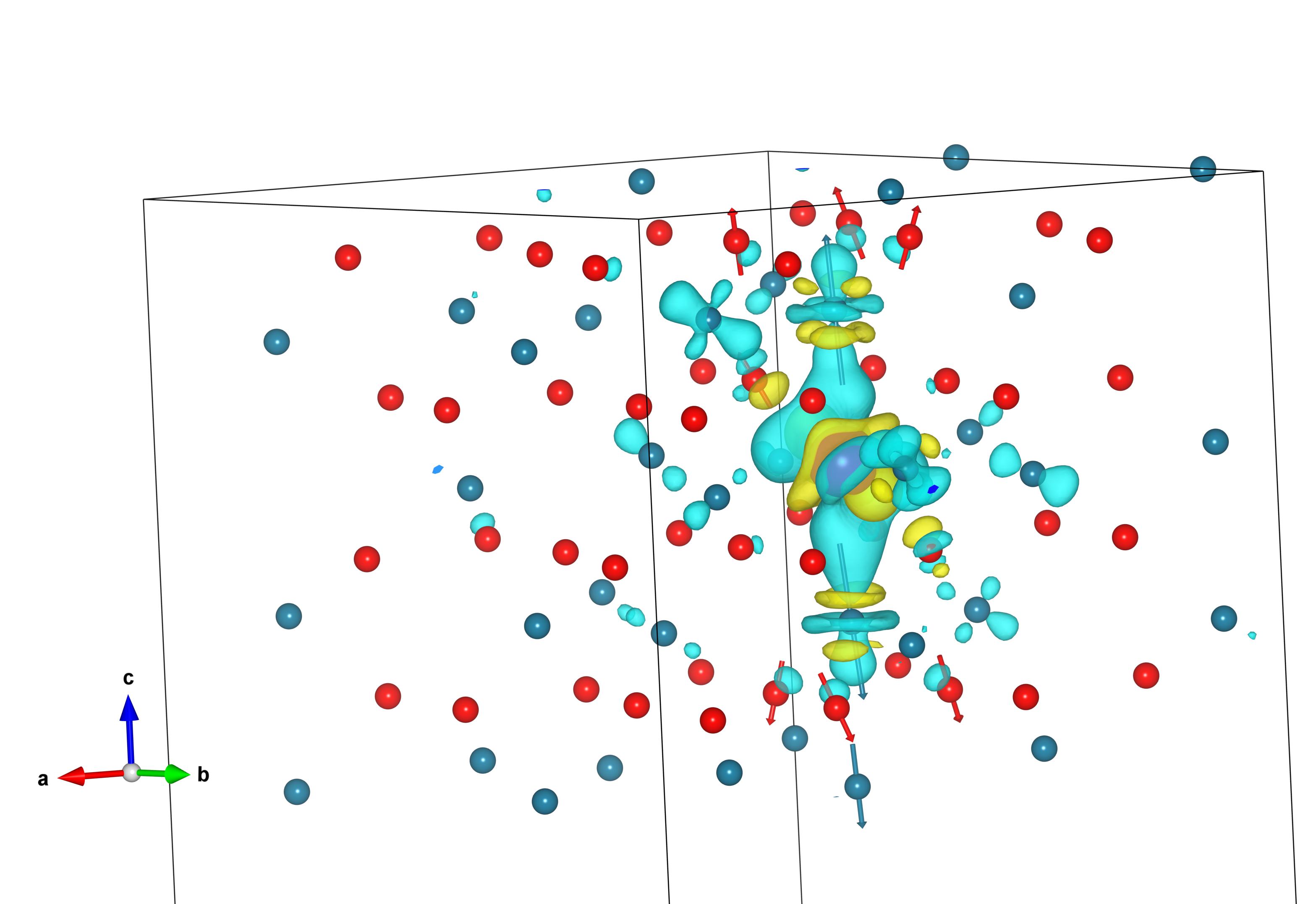}
    \caption{${Cr}_i^{+2}$}
    \label{fig:Crinterdispvectq2}
  \end{subfigure}%
  \begin{subfigure}{.5\textwidth}
    \centering
    \includegraphics[width=1\linewidth]{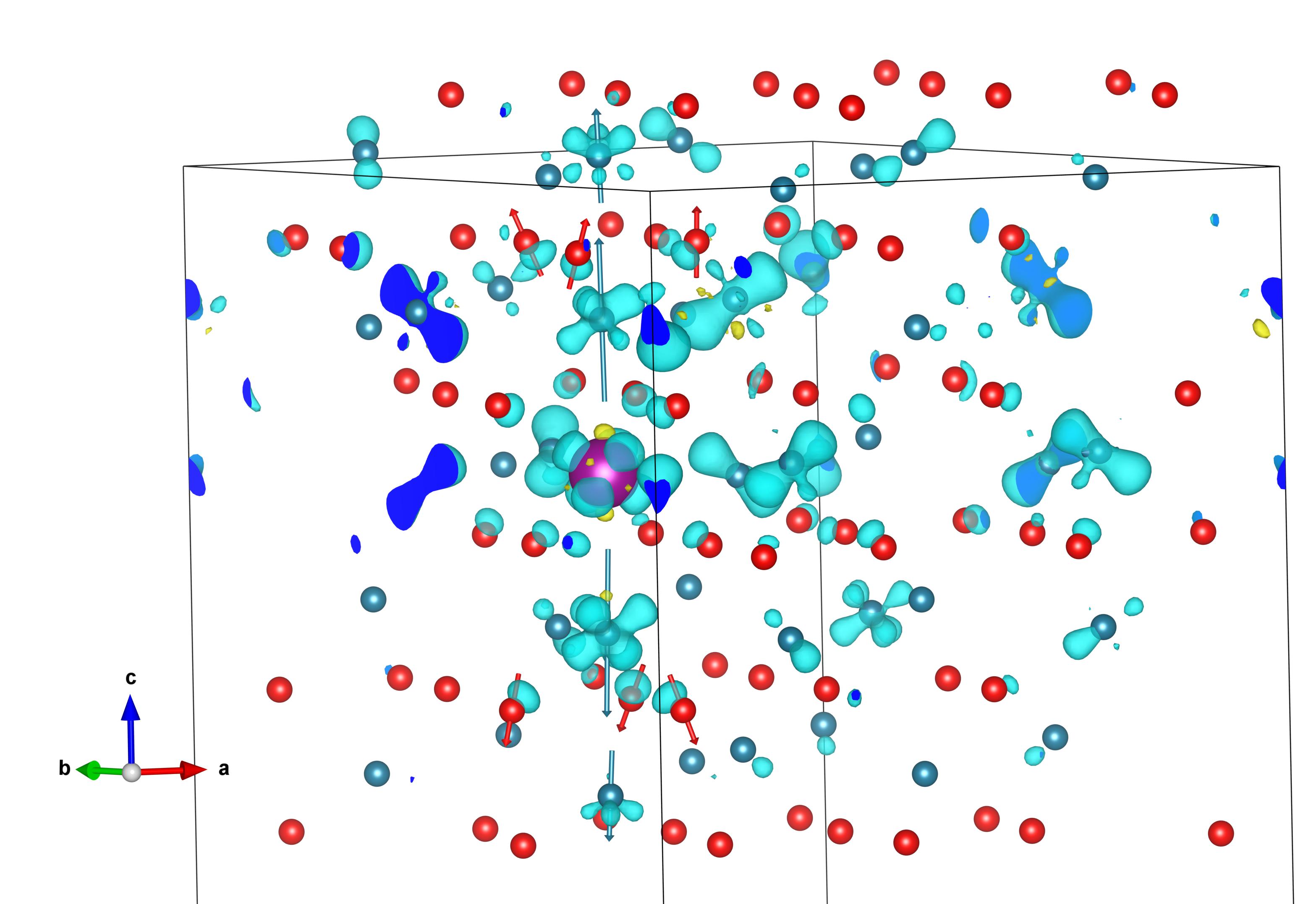}
    \caption{${Cr}_i^{+3}$}
    \label{fig:Crinterdispvectq3}
  \end{subfigure}
  \caption{Relaxed geometry surrounding the Cr 
  interstitial (denoted with magenta sphere)  for 
  different charge states. For non-zero charge states, the 
  density of the hole charge is also plotted. The teal and 
  red arrows indicate the displacement of Cr and O ions 
  respectively. Yellow and blue isosurfaces indicate the 
  excess or deficit electronic charge at 
  5 millielectrons/$\AA^3$.}
\label{fig:Crinterdispvect}
\end{figure}

In addition to PyCDT and pymatgen, we utilized 
phonopy~\cite{phonopy}, and VESTA~\cite{vesta} for generating inputs to 
DFT calculations, post-processing, and plotting the
DFT calculations results. 

\section{Results and Discussion \label{sec:results}}
\begin{wrapfigure}{r}{0.5\linewidth}
    \vspace{-10pt}
  \begin{subfigure}{\linewidth}
    \centering
    \includegraphics[width=1\linewidth]{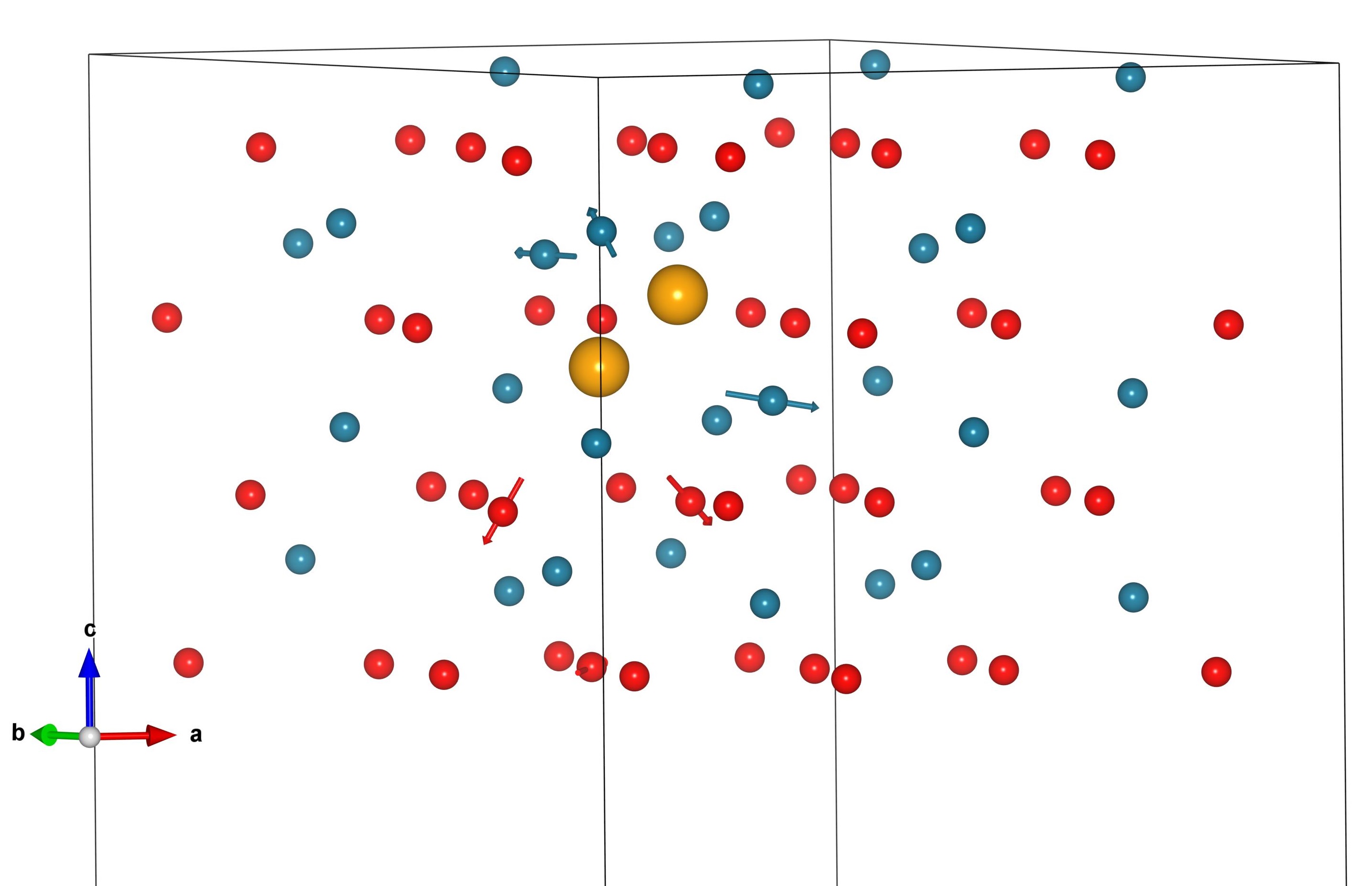}
    \caption{${O}_i^0$}
    \label{fig:Ointerdispvectq0}
  \end{subfigure}\\
  \begin{subfigure}{\linewidth}
    \centering
    \includegraphics[width=1\linewidth]{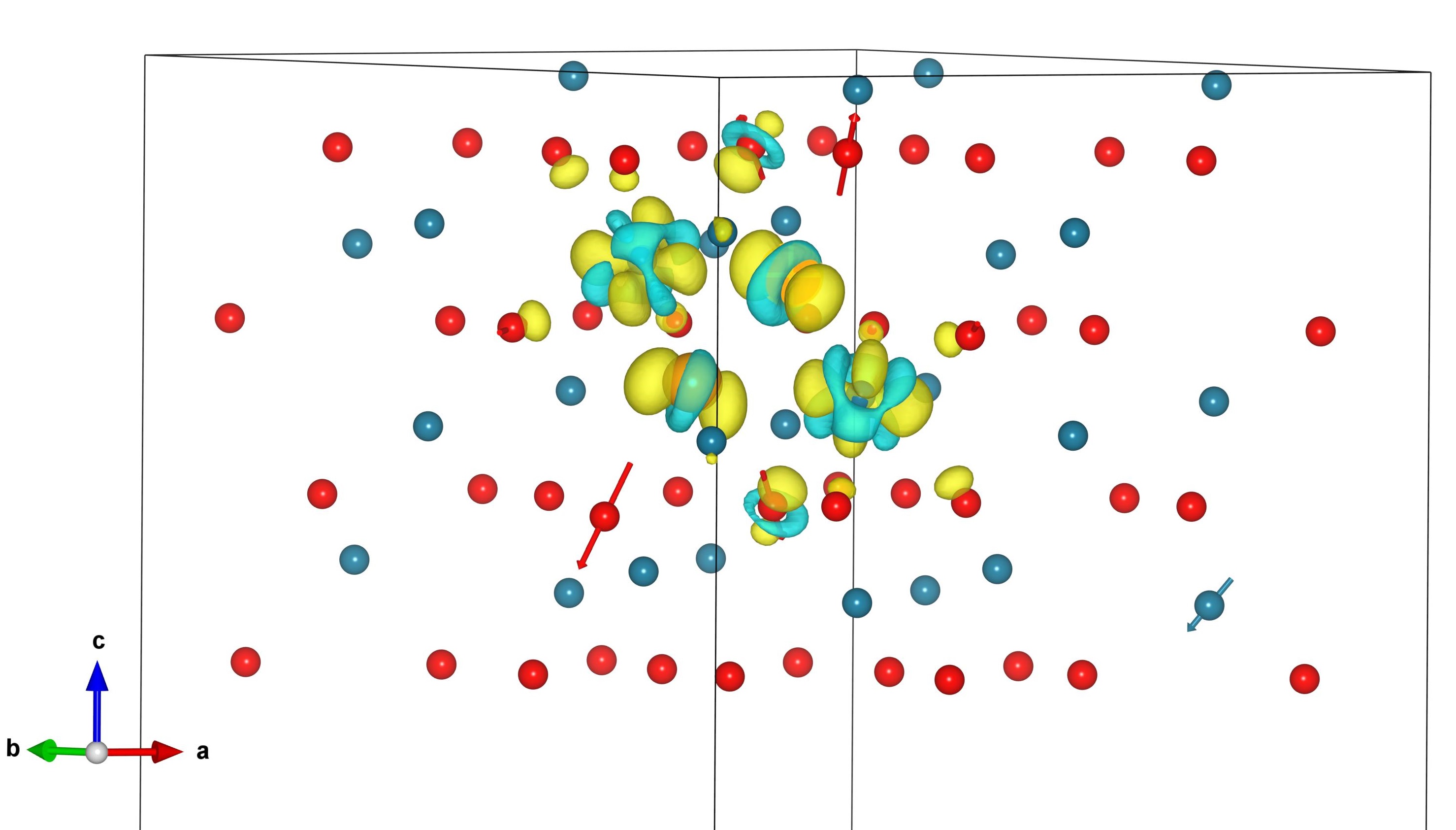}
    \caption{${O}_i^{-1}$}
    \label{fig:Ointerdispvectq1n}
  \end{subfigure}\\
  \begin{subfigure}{\linewidth}
    \centering
    \includegraphics[width=1\linewidth]{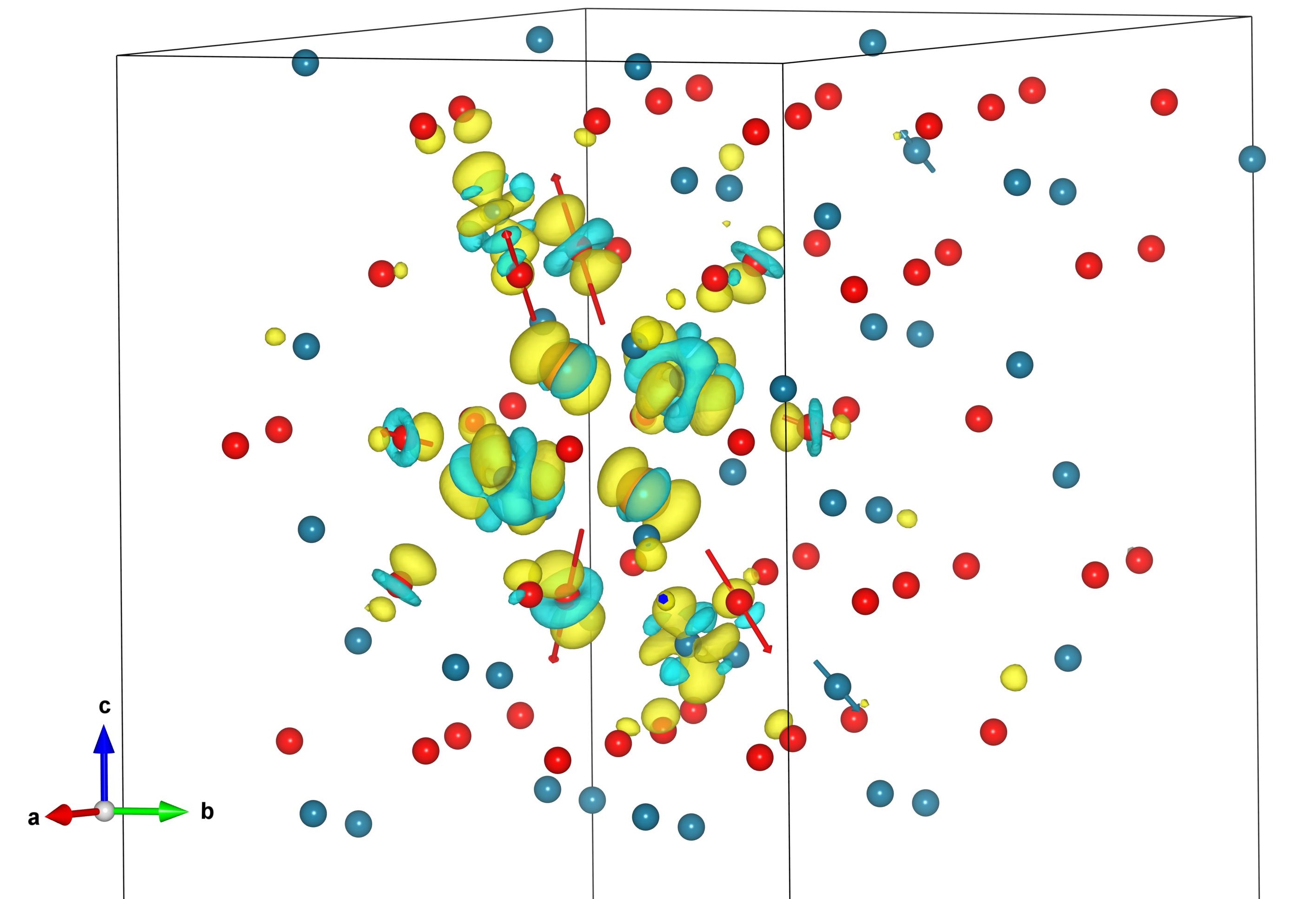}
    \caption{${O}_i^{-2}$}
    \label{fig:Ointerdispvectq2n}
  \end{subfigure}
    \vspace{-10pt}
  \caption{Relaxed geometry surrounding the O dumbbell 
  interstitial (denoted with orange spheres) for different 
  vacancy charge states. For charged $O_i$, the 
  distribution of excess electrons is also plotted.
  The teal and red spheres and the corresponding arrows 
  indicate the Cr and O ions and their displacements 
  respectively. Orange spheres indicate the dumbbell 
  interstitial. Yellow and blue isosurfaces indicate the 
  excess or deficit electronic charge at 5 millielectrons/$\AA^3$.}
\label{fig:Ointerdispvect}
    \vspace{-10pt}
\end{wrapfigure}
\subsection{Potential Interstitial Sites}
The order-parameter based interstitial finding algorithm introduced by 
Zimmerman et al.~\cite{zimmermann2016}  identified  five potential interstitial sites and
their positions  are shown in Figure~\ref{fig:interstitialsites}. 
Among the potential interstitial sites, three are tetrahedral sites and the remaining two are octahedral sites, 
which are defined by their coordination numbers as well as the coordinating elements. 
$O_{i,m-Cr,n-O}$ represents an octahedral interstitial site surrounded by $m$ Cr 
sites and $n$ O sites. Similarly $T_{i, m'-Cr,n'-O}$ represents a tetrahedral 
interstitial site with $m'$ Cr sites and $n'$ O sites as neighbors.  Of 
these sites, the $T_{i, 3-Cr,1-O}$ site was found to be too close to the 
regular corundum lattice sites and ignored. The remaining four sites were 
populated either by Cr or O and the structure of the resultant defect 
supercells were optimized.   
The Cr and O interstitial defects were initially assigned 
with  charge  states in the range of [0,6] and [-2, 0] 
respectively.

The resulting uncorrected
interstitial formation energies are shown 
in Figure~S1 in SI.
The large differences in the calculated formation energies for the different interstitial positions in the lattice indicate that $O_{i,6-O}$  site is 
the most favorable interstitial site for both O and Cr interstitials (denoted as 
$O_{i_{oct,O6}}$ and $Cr_{i_{oct,O6}}$ respectively). Hence for
further analysis, we considered only the 6-O coordinated octahedral
interstitial site. Further on, we use standard notation of  $Cr_i$ and $O_i$ 
for Cr and O interstitials at the 6-O coordinated octahedral site instead of
$Cr_{i_{oct,O6}}$ and $O_{i_{oct,O6}}$,  respectively. 
The uncorrected formation 
energies $Cr_i$  for charge states of  4, 5 and 6 are very high when compared 
to those of the charge states in the range [0, 3], and hence the charge states 
of 4$-$6 for $Cr_i$ are ignored in this study.

\subsection{Interstitials}
The relaxed interstitial structures shown in Figure~\ref{fig:Crinterdispvect} 
 indicate that Cr interstitials occupy the unoccupied 
octahedral sites in the Cr bilayer coordinated by the O sublattice.
In contrast, the O interstitials shown in Figure~\ref{fig:Ointerdispvect} 
relax to a dumbbell configuration with each 
of the O ion in the dumbbell lobes residing at the unoccupied octahedral sites. 

The effect of interstitial charge on the displacement of 
the lattice ions around the interstitial is  shown by the ion displacement vectors in 
Figures~\ref{fig:Crinterdispvect} and 
\ref{fig:Ointerdispvect}  and  the magnitudes of ionic displacements are
summarized in Tables~S1 and S2 in SI for Cr and O
interstitials, respectively.
The relaxation of the ions surrounding the interstitials was found to be  anisotropic.
The nearest Cr ions to a Cr interstitial shift away from the interstitial 
with the maximum displacement at 0.3 \AA. The
amount of displacement was found to be dependent on the interstitial charge and the 
distance of the ions from the interstitial. In particular, the nearest O ions 
have an outward displacement of up to 0.22 \AA{} except in the case of  $Cr_i^3$, 
where the nearest O ions move towards the Cr interstitial by  0.12 -- 0.15\AA. 
For neutral O interstitials, only  two Cr ions that are at 1.9~\AA{} distance move away 
from the  interstitial. For charged O interstitials, O ions at 2.7 and 
2.9~\AA{} distance are displaced by 0.36~\AA{} and 0.23~\AA{} respectively 
for -2 charge, and undergo even less shifting for q $= -1$. The relatively 
small magnitude of displacements and the small number of ions with any 
noticeable displacement indicate that interstitials do not introduce any 
significant local distortion in the lattice of \ce{Cr2O3}.
\begin{figure*}
  \begin{subfigure}{.5\textwidth}
    \centering
    \includegraphics[width=1\linewidth]{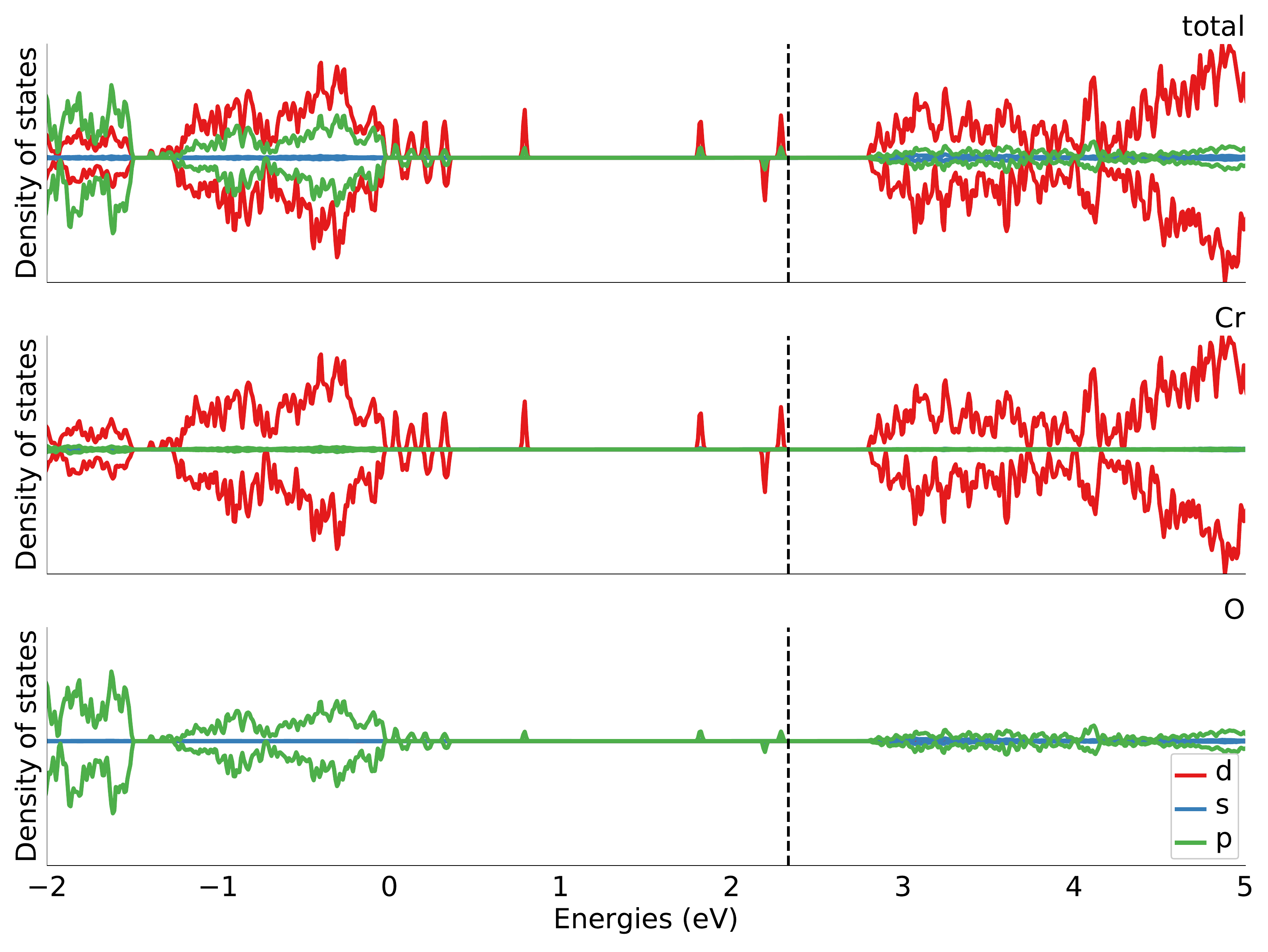}
    \caption{${Cr}_i^0$}
    \label{fig:Crinterdosq0}
  \end{subfigure}%
  \begin{subfigure}{.5\textwidth}
    \centering
    \includegraphics[width=1\linewidth]{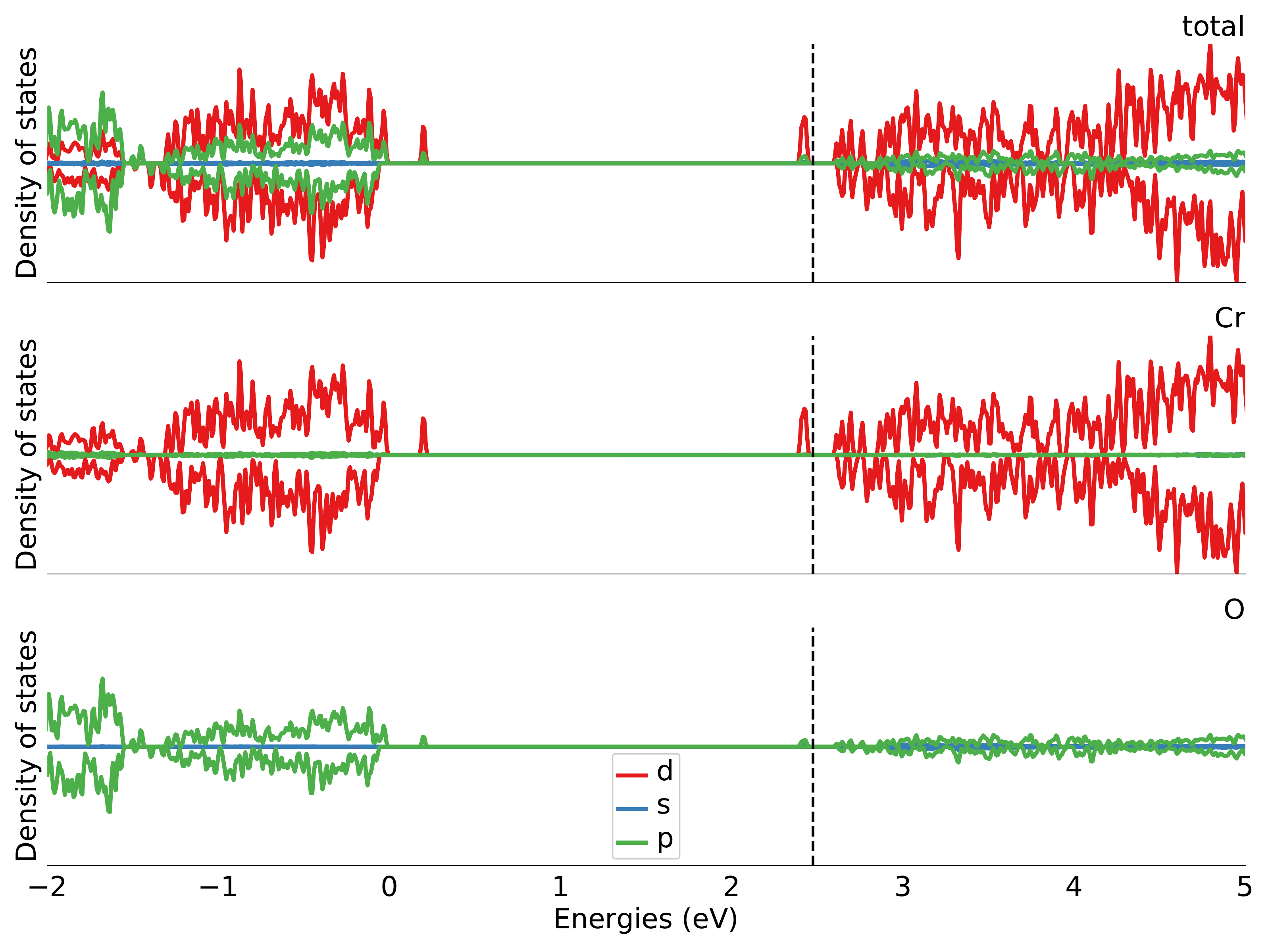}
    \caption{${Cr}_i^{1}$}
    \label{fig:Crinterdosq1}
  \end{subfigure}
  \begin{subfigure}{.5\textwidth}
    \centering
    \includegraphics[width=1\linewidth]{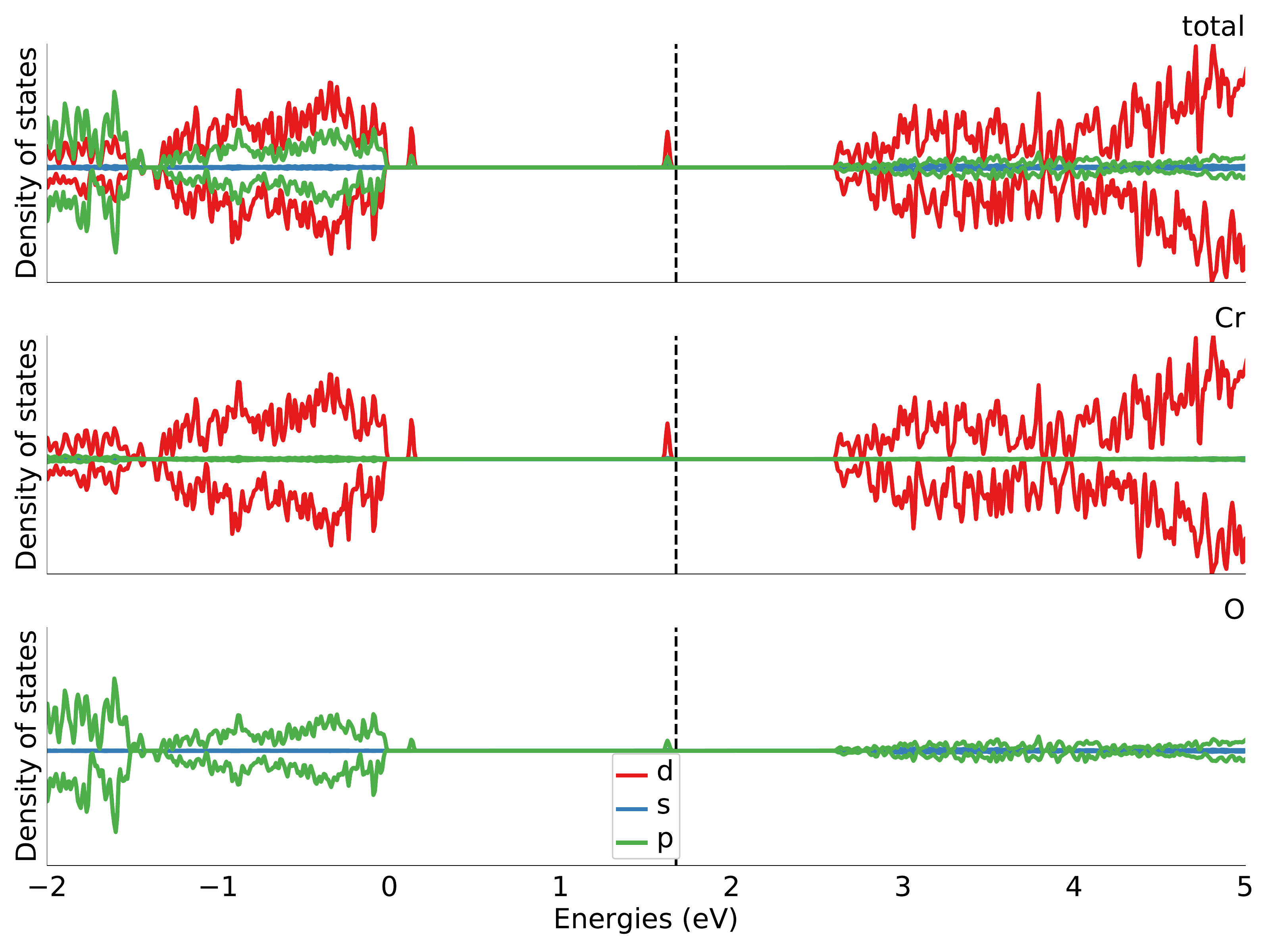}
    \caption{${Cr}_i^{2}$}
    \label{fig:Crinterdosq2}
  \end{subfigure}%
  \begin{subfigure}{.5\textwidth}
    \centering
    \includegraphics[width=1\linewidth]{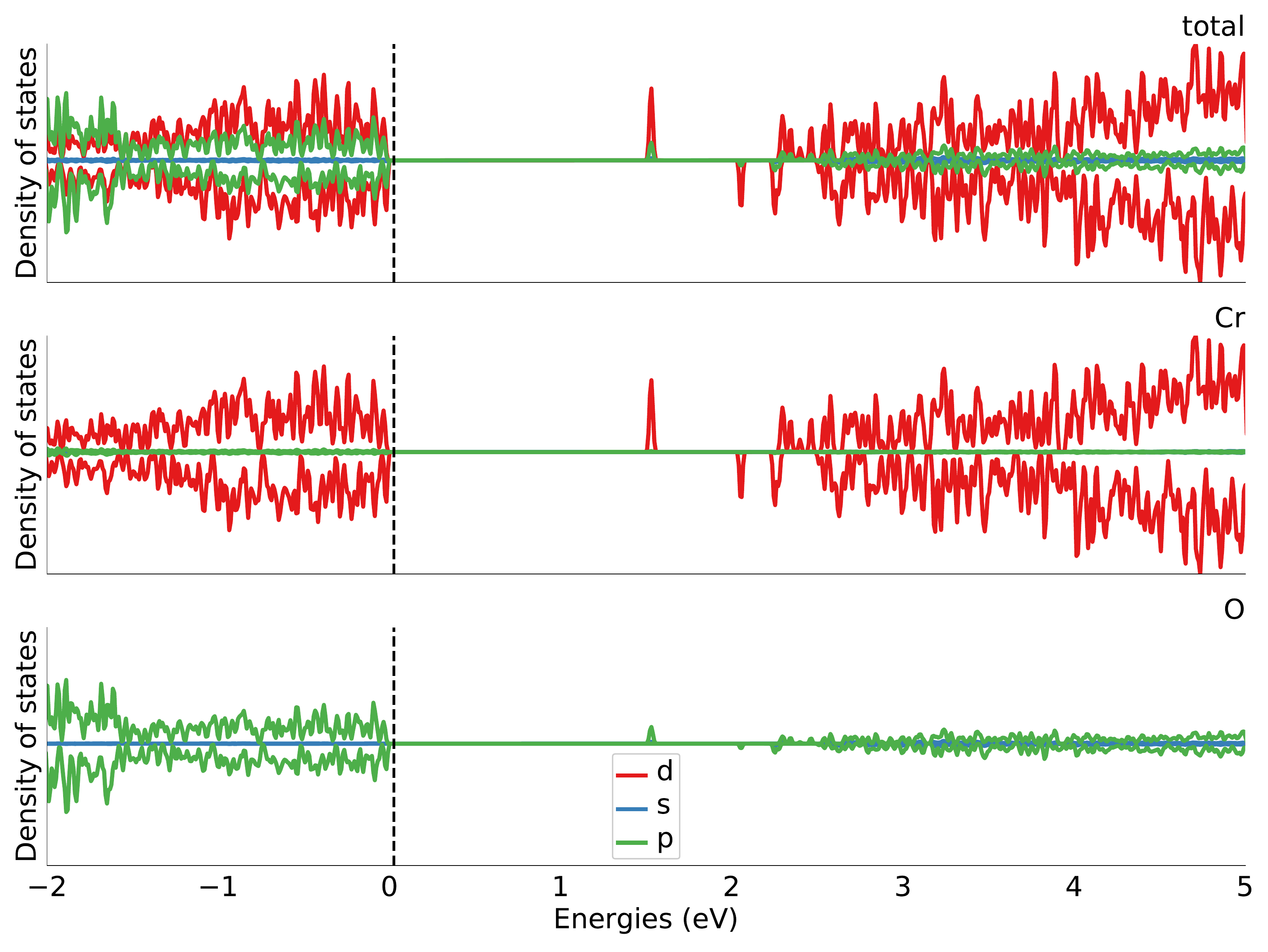}
    \caption{${Cr}_i^{3}$}
    \label{fig:Crinterdosq3}
  \end{subfigure}
\caption{Element projected orbital density of states averaged over all sites for Cr interstitials in different charge states. Dashed vertical line marks the position of the Fermi level.}
\label{fig:Crinterdos}
\end{figure*}

To evaluate the localization of holes in charged $Cr_i$ and 
electrons in charged $O_i$, the difference between charge distribution in 
a charged interstitial  and the charge distribution of the corresponding 
neutral interstitial with identical structure is plotted in 
Figures~\ref{fig:Crinterdispvect} and  \ref{fig:Ointerdispvect}.
In some of the previous studies of defects in oxides with excess holes, 
a non-zero Hubbard correction in the range of 4.0 $-$ 5.5 eV was applied for O-2p orbitals to properly
account for the hole charge distribution\cite{huang2014,kehoe2016,Carey2016}. 
However, in undoped \ce{Cr2O3}, 
applying $U_{O_{2p}} = 5$,  did not produce any noticeable difference in
the hole charge distribution (see Figure S2 in SI).
For $Cr_i^{1}$, the hole charge shown in Figure~\ref{fig:Crinterdispvectq1} is mainly localized on two lattice Cr ions closest to
the interstitials located at a distance of 2.45 \AA{} along the c-axis and 
2.99 \AA{} in the basal plane from the interstitial. 
The major portion of the remaining hole charge is localized on two Cr ions in 
the basal plane at a 
distance of 2.94 \AA{} from the interstitial. In the case of $Cr_i^{2}$,  the majority of the 
2-hole charge is localized on the Cr interstitial itself. Some of the hole charge is
distributed on the two nearest Cr ions along the c-axis at 2.4~\AA, two Cr ions in the basal plane at
2.94~\AA{} and two Cr ions at 3.8~\AA{} along the [221] direction. For $Cr_i^{3}$, the 3-hole charge

\begin{wrapfigure}{r}{0.45\textwidth}
    \vspace{-20pt}
  \begin{subfigure}{\linewidth}
    \centering
    \includegraphics[width=1\linewidth]{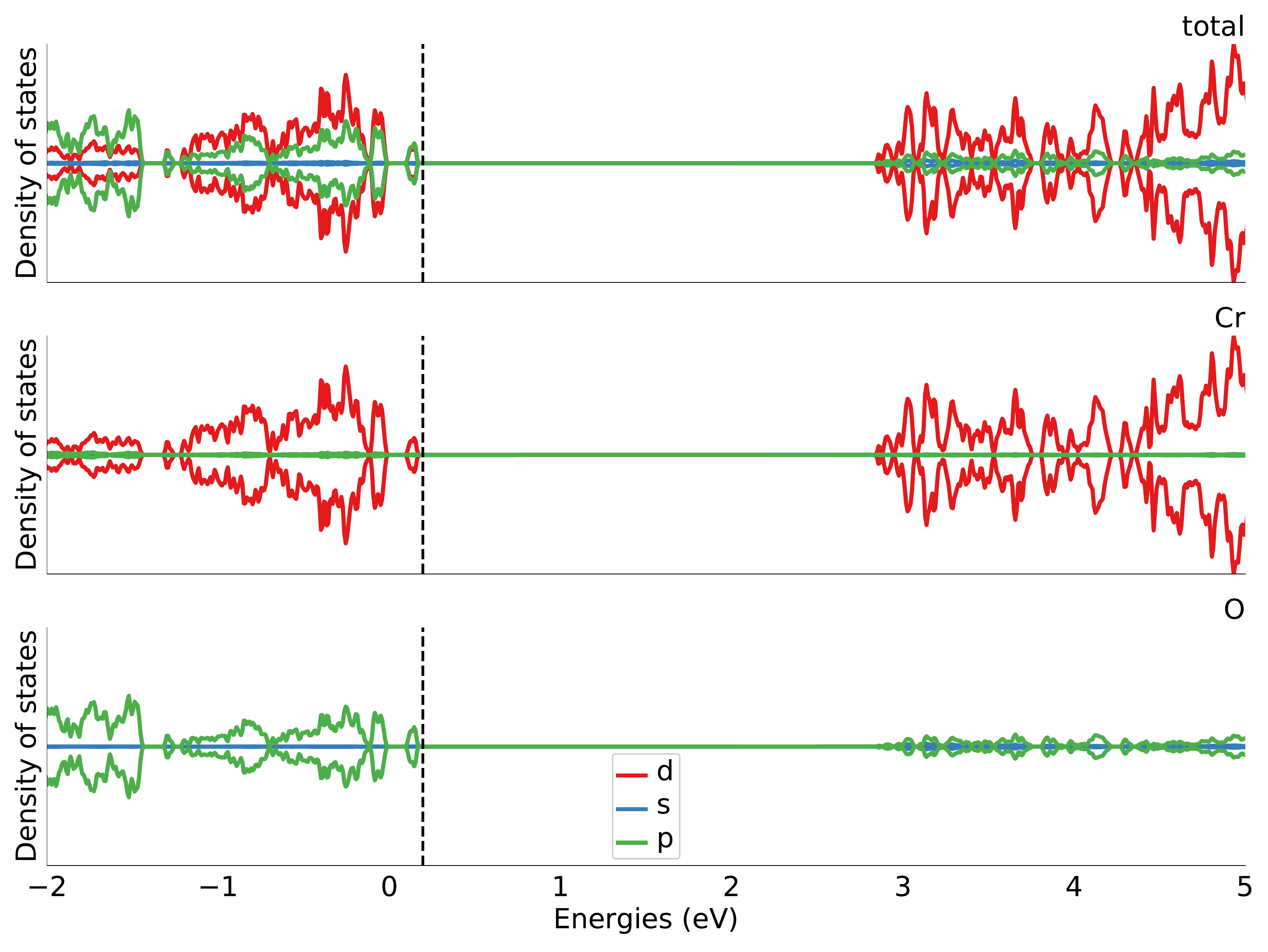}
    \caption{${O}_i^0$}
    \label{fig:Ointerdosq0}
  \end{subfigure}\\
  \begin{subfigure}{\linewidth}
    \centering
    \includegraphics[width=1\linewidth]{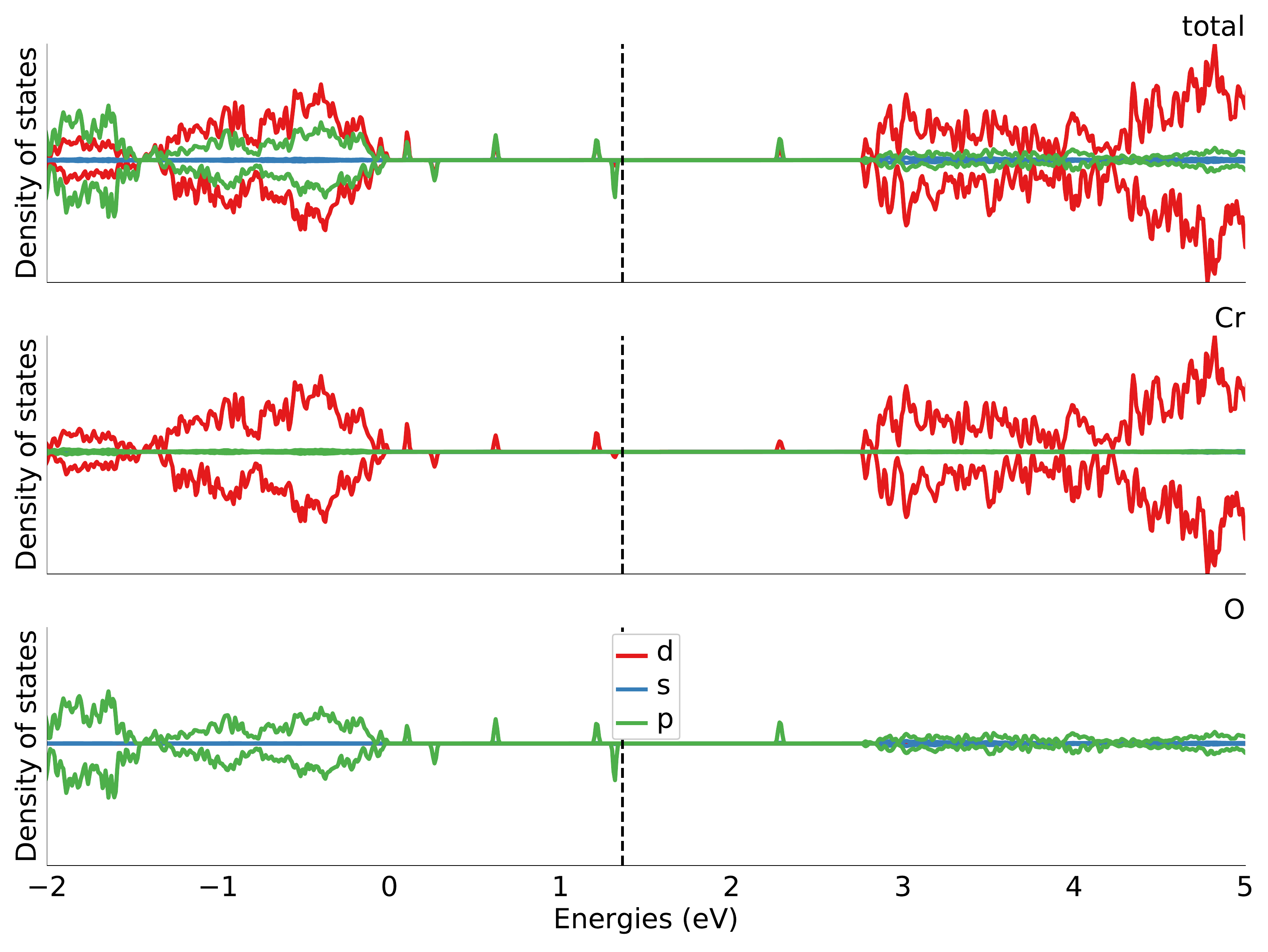}
    \caption{${O}_i^{-1}$}
    \label{fig:Ointerdosq1n}
  \end{subfigure}\\
  \begin{subfigure}{\linewidth}
    \centering
    \includegraphics[width=1\linewidth]{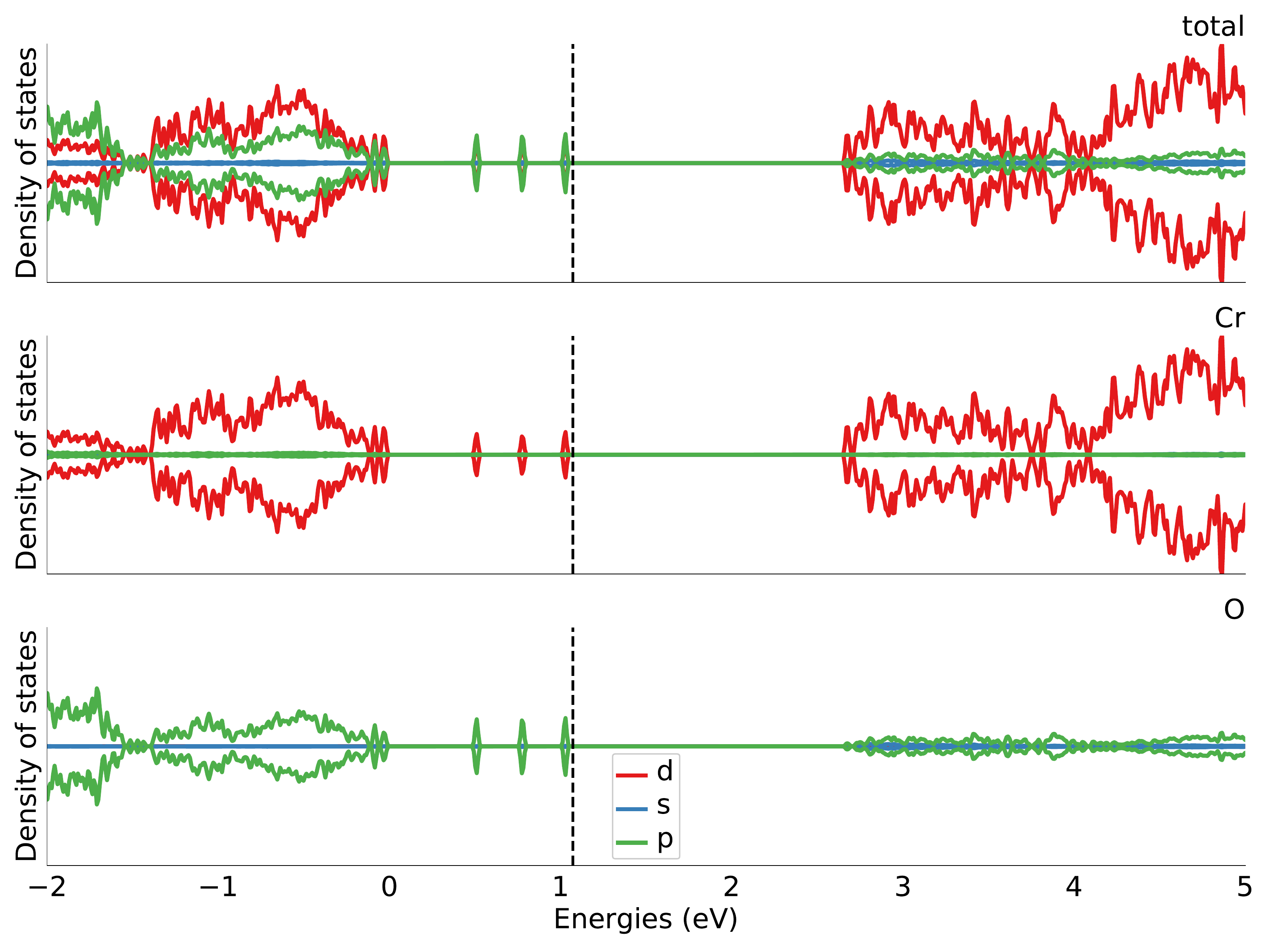}
    \caption{${O}_i^{-2}$}
    \label{fig:Ointerdosq2n}
  \end{subfigure}
    \vspace{-10pt}
\caption{Element projected orbital density of states averaged over all sites for O interstitials. Dashed vertical line marks the position of the Fermi level.}
\label{fig:Ointerdos}
    \vspace{-10pt}
\end{wrapfigure}
The charge distribution plots of excess holes indicate that the holes are neither
completely localized on individual ions nor completely delocalized. This 
result is consistent with the predictions of Lany~\cite{lany2015} and 
Kehoe et al~\cite{kehoe2016}. According to Lany, the non-bonding Cr-3d $t_{2g}$ orbitals 
at the top of the valence band are 
responsible for the delocalized nature of the holes.
The holes in \ce{Cr2O3} are predicted to be band conducting with heavy effective 
mass~\cite{lany2015}.

The excess electron density of $O_i^{-1}$ is localized 
mainly on the O ions forming the lobes of a dumbbell 
interstitial and the two nearest Cr ions. For $O_i^{-2}$, the excess 
2-electron charge  is also distributed over the adjacent 
Cr and the O ions in the basal planes that are at 
2.4 $-$ 2.5~\AA{} distance from either of the 
interstitial O ions.

The electronic density of states (DOS) plots of the optimized Cr and O 
interstitial structures are presented in Figures~\ref{fig:Crinterdos} 
and \ref{fig:Ointerdos}, respectively. The computed bandgap in these figures is 
2.8 eV~\cite{medasani2017}. The plots reveal that the energy 
states corresponding to both Cr and O interstitials are located in 
the bandgap, and are of hybrid O-$p$ and Cr-$d$ character, which is 
similar to the nature of the vacancy states~\cite{medasani2017}. The defect levels of the 
neutral Cr interstitial, which are located in both the lower and upper 
parts of the bandgap, are occupied with electrons and the Fermi level 
is closer to the conduction band. For positively charged Cr interstitials, the 
unoccupied defect levels are shifted up on the energy scale. The upward shift 
of the empty interstitial levels increases with the 
charge of Cr interstitial. This effect can  be  attributed to the reduction 
in their screening due to the loss of electrons. For $Cr_i^3$,  the
majority of the defect levels are empty and are shifted closer 
of  the conduction band. 

The DOS plots of $O_i$ indicate that occupied $O_i^0$ levels are located at 
the bottom of the bandgap region near the valence band maximum (VBM) (Figure~\ref{fig:Ointerdos}). 
The unoccupied $O_i^0$ levels are in resonance with the conduction 
band. When the $O_i$ interstitial charge is increased negatively, the interstitial gains electrons
and the newly occupied defect levels shift down towards 
the center of the bandgap. For $O_i^{-2}$, the occupied defect levels 
reside in the lower half of the bandgap. Interestingly, the highest 
occupied level of $O_i^{-1}$ is higher than that of the $O_i^{-2}$ and is characterized by
spin splitting of the defect levels. Such relative position of defect levels can be attributed to the 
relatively lower screening experienced by highest occupied defect level  
in $O_i^{-1}$ compared to that of $O_i^{-2}$.

\begin{wrapfigure}{r}{0.6\textwidth}
  \begin{subfigure}{.3\textwidth}
    \centering
    \includegraphics[width=1\linewidth]{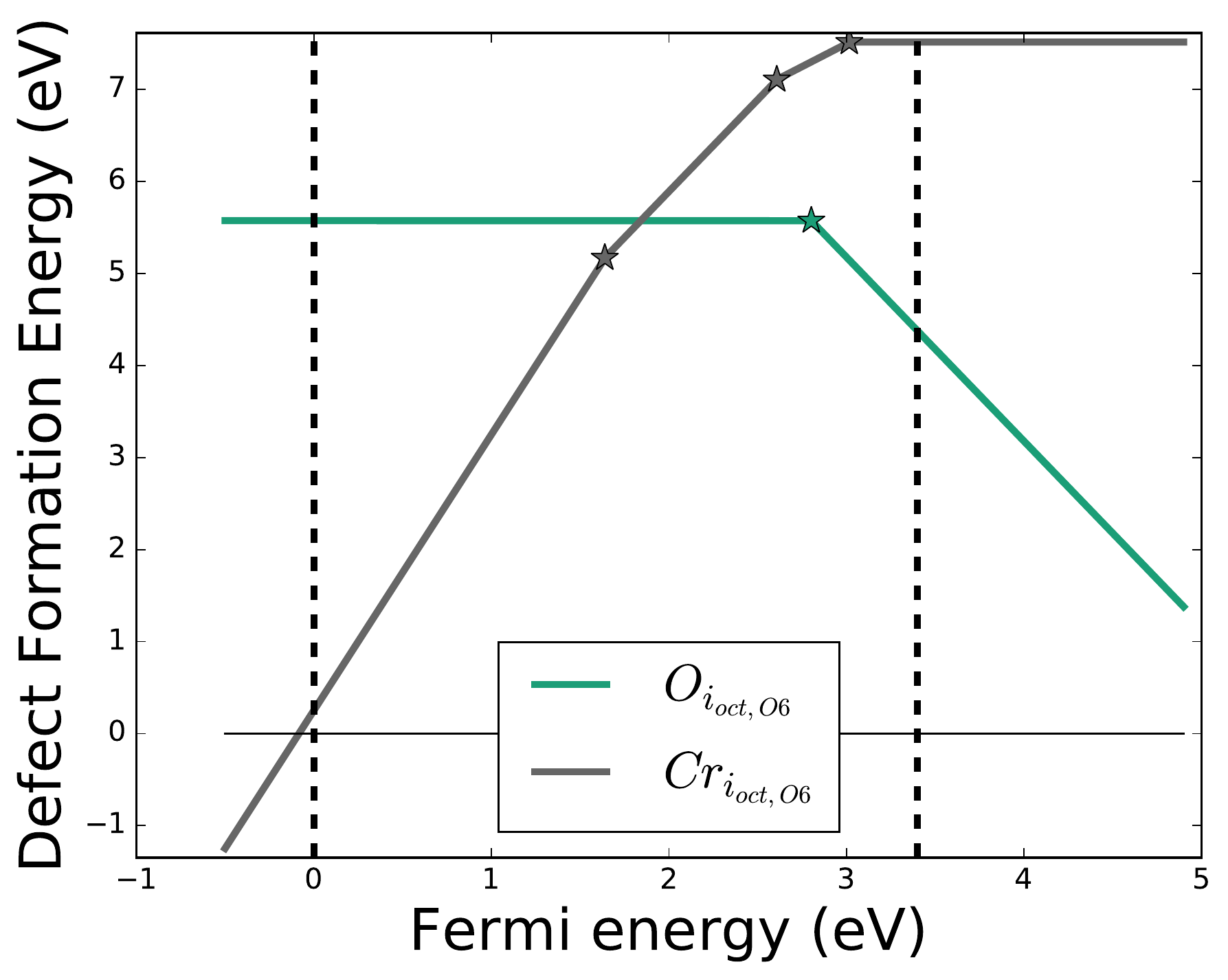}
    \caption{Cr rich boundary}
    \label{fig:interformenqbgcorrCrrich}
  \end{subfigure}%
  \begin{subfigure}{.3\textwidth}
    \centering
    \includegraphics[width=1\linewidth]{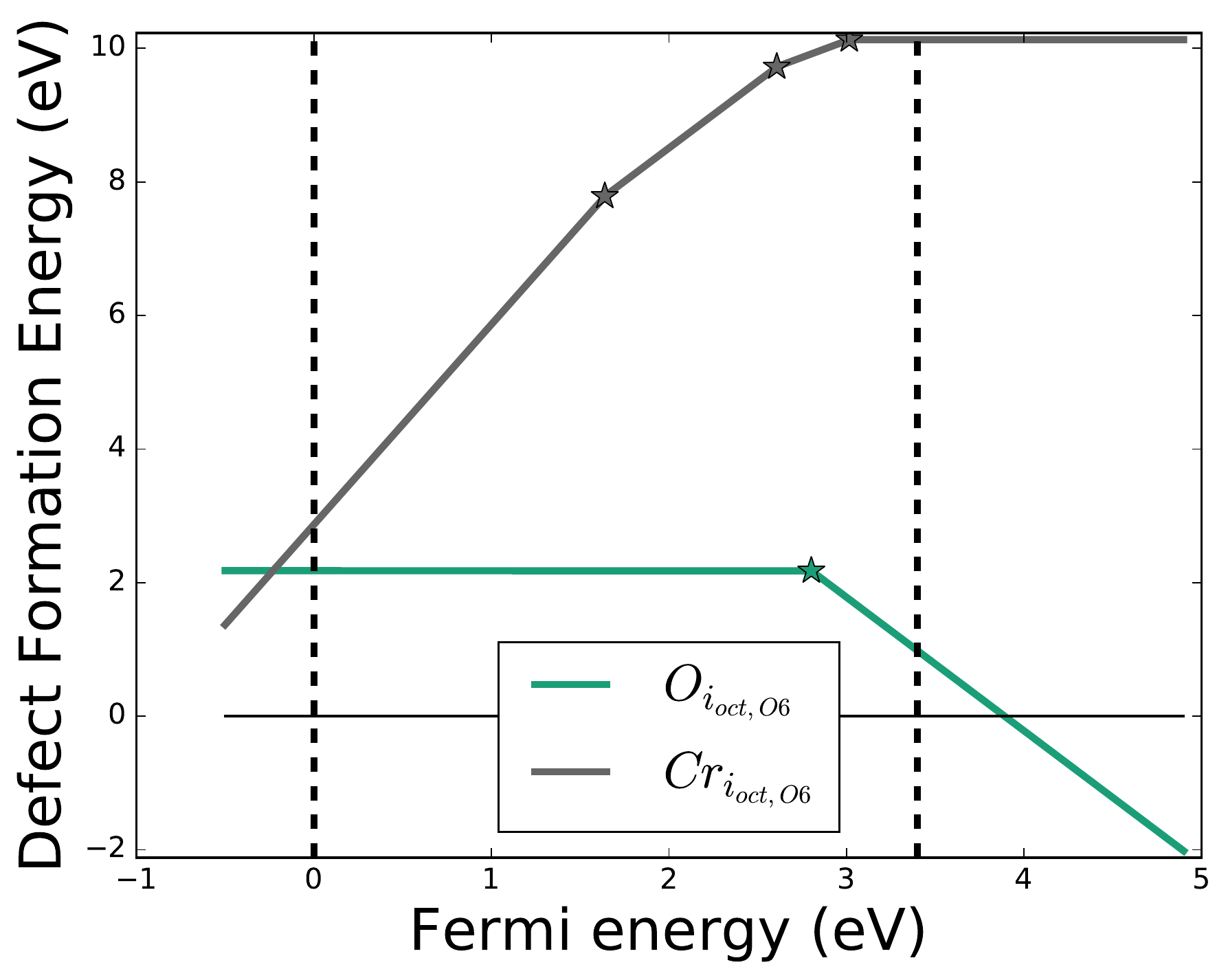}
    \caption{O rich boundary}
    \label{fig:interformenqbgcorrOrich}
  \end{subfigure}%
  \caption{Formation energies of Cr and O interstitials  after applying 
  electrostatic and bandgap related  corrections. Black solid line corresponds 
  to $E_f({Cr}_i)$ and green dashed line refers to $E_f({O}_i)$. The dashed
  vertical lines denote the edges of bandgap region with 0-level representing
  valence band maximum (VBM).}
\label{fig:interformenqbgcorr}
\end{wrapfigure}

\begin{wraptable}{r}{0.6\textwidth}
    \centering
    \caption{Interstitial formation energies (in eV) of \ce{Cr2O3} with Fermi level ($E_F$) at 0.1 eV above VBM, 0.1 eV below CBM, and in the middle of the bandgap.}
\begin{tabular}{lrccccccc}
    \hline
    Defect &  q & \multicolumn{6}{c}{Interstitial Formation Energy} \\ 
    ($E_F=$)        &    & \multicolumn{2}{c}{VBM+0.1 eV} & \multicolumn{2}{c}{CBM-0.1 eV} & \multicolumn{2}{c}{VBM+$E_g$/2}  \\
                &    &  Cr rich & O rich & Cr rich & O rich  &  Cr rich & O rich  \\
    \hline
    \multirow{3}{*}{$O_i$}    &  0 & 5.57 & 2.18 & 5.57 & 2.18 & 5.57 & 2.18 \\
                              & -1 & 8.69 & 5.29 & 5.49 & 2.09 & 7.09 & 3.69 \\
                              & -2 &10.98 & 7.58 & 4.58 & 1.18 & 7.78 & 4.38 \\
    \hline
    \multirow{4}{*}{$Cr_i$}   & 0 & 7.51 &10.13 & 7.51 &10.13 & 7.51 &10.13 \\
                              & 1 & 4.60 & 7.21 & 7.80 &10.42 & 6.20 & 8.82 \\
                              & 2 & 2.09 & 4.71 & 8.49 &11.11 & 5.29 & 7.91 \\
                              & 3 & 0.55 & 3.17 &10.15 &12.77 & 5.35 & 7.97 \\
    \hline
\end{tabular}
\label{tab:interen}
\end{wraptable}
The interstitial formation energies were corrected for 
electrostatic interactions between periodic images of charged simulation cells and for the 
under-predicted DFT bandgap.  
Electrostatic corrections were obtained using anisotropic FNV (Freysoldt,
Neugebauer, and Vande Walle) method~\cite{freysoldt2009,kumagai2014} 
(see   Table S3 in SI for details). 
The transition levels of Cr and O interstitials  with respect to the 
VBM were 
corrected for the DFT bandgap error using the scheme proposed  by Janotti 
and Vande Walle\cite{janotti2007}. The resulting corrections to the transition
levels and the associated corrections to the interstitial formation 
energies are summarized in SI Tables S4 and S5, respectively. The resultant formation energies 
were plotted with respect to the position of the Fermi level in Figure~\ref{fig:interformenqbgcorr} for both 
O-rich and Cr-rich conditions. These conditions were
defined by the upper and lower limits of Cr and O chemical potentials 
determined in our previous study~\cite{medasani2017}. Here, we 
focused on the 0 K DFT computed phase stability, and the 
investigation of the effect of $p_{O_2}$ and temperature on interstitial 
formation energies is deferred to  a future study.

The formation energies reveal that for Cr-rich and p-type conditions, as the 
Fermi level is in the lower half of the
bandgap, Cr interstitials are energetically more 
favorable compared to O interstitials. Under n-doped conditions, corresponding to 
the Fermi level being in the upper half of bandgap, O interstitials 
become the dominant interstitial defects  even in Cr rich conditions (Figure~\ref{fig:interformenqbgcorr}). 
However, under n-doped, Cr-rich conditions, both $V_{Cr}$ and $V_O$ have 
lower formation energies than interstitials. For O rich conditions, O interstitials are energetically more favorable over 
Cr interstitials under all doping conditions. The transition levels of both 
Cr and O interstitials, which are denoted as stars in the 
Figure~\ref{fig:interformenqbgcorr} (see also the last 
column of SI Table S4), are deep
in the bandgap. While the $Cr_i$ +3/+2 transition level is very deep and
1.64~eV  above VBM, the other transition levels are relatively shallow
and are less then 0.6 eV below the conduction band minimum (CBM). 
The predicted transition levels for O interstitials (last
column of SI Table S4) indicate that 
either of neutral or -2 charge states has lower formation energy 
when compared to -1 charge state at any Fermi level in the bandgap.  The 
phonon DOS plot in  SI Figure S4 (a) shows that $O_i^{-1}$ is a stable defect 
with no soft modes. This suggests that the concentration of $O_i^{-1}$ 
would be relatively insignificant under any physical condition.

Our grand canonical calculations of defect chemistry at finite 
temperatures (Figure S3 and S4 in SI) indicate that undoped \ce{Cr2O3} is a stable 
semiconductor with the Fermi level located in the center region of the band gap. 
Our calculations predict that at high temperatures ($>$~800~K), 
intrinsic \ce{Cr2O3} 
is weakly \textit{n}-type and  at low (300~$-$~400 K) and intermediate 
(400~$-$~800 K) temperatures, 
intrinsic \ce{Cr2O3} is slightly \textit{p}-type.
When \ce{Cr2O3} acts as a protecting layer on a metal substrate,
the metal-oxide boundary has effectively very low $p_{O_2}$. At 
low $p_{O_2}$, the formation energies of $V_O$ and $V_{Cr}$ 
at $E_F=1.7$~eV, are equal to 2.6 eV and 4.0 eV 
respectively~\cite{medasani2017}, and
are lower than those of $Cr_i$ and $O_i$, which are equal to 5.3 and 5.6 eV, respectively. 
This indicates that at low $p_{O_2}$, $V_O$ could be the dominant defect in
undoped \ce{Cr2O3} followed by $V_{Cr}$. This contradicts the conclusions from some of the experimental studies on self-diffusion in 
\ce{Cr2O3}~\cite{latanision1967,Schmucker2016}, 
where $V_O$ and $Cr_i$ were suggested as the dominant defects in 
\ce{Cr2O3} at very low $p_{O_2}$. 
It is noteworthy, that
\ce{Cr2O3} passivation films on Ni alloy substrate could be  p-type 
doped due to trace amounts of 
Ni substitution~\cite{Carey2016}. For Ni doping, Fermi level  in 
\ce{Cr2O3} stabilizes at 0.4 eV above VBM~\cite{Carey2016}.
Under such conditions, 
$Cr_i$ defects have lower formation energy compared to $V_{Cr}$ and both
$Cr_i$ and $V_{O}$ have formation energies approximately equal to 1.5 eV, 
making $Cr_i$ and $V_O$ the dominant defects. 
Similarly, bulk \ce{Cr2O3} samples can be 
strongly p-type doped with $E_F$ close to VBM, when 
divalent ions such as Mg are present as 
impurities~\cite{Carey2016}. Under such conditions, 
our calculations suggest that $Cr_i$ could be the dominant defect 
in bulk \ce{Cr2O3} in place of $V_{Cr}$. Based on these results, 
the \ce{Cr2O3} samples used Ref.\citenum{latanision1967} and 
\citenum{Schmucker2016} 
could be inadvertently \textit{p}-type doped due to trace impurities. 
Furthermore, the presence of 
either an additional spinel layer or hydrogenated 
water  opposite to the metal-oxide side could result in reducing 
conditions for the  entire \ce{Cr2O3} layer in the PWR environments.
In such conditions, the  dominant defects could be either Cr 
interstitials or Cr vacancies depending on the Fermi level.

With vacancies included, under heavily oxidizing conditions, $O_i$ and $V_{Cr}$ are the 
dominant defects. When the Fermi level is above 2.2 eV, \ce{Cr2O3} is 
unstable against $V_{Cr}$\cite{medasani2017} indicating that 
compensating Cr vacancies will form if \ce{Cr2O3} is heavily doped with 
electron donating species. Under \textit{p}-doped conditions, $V_{Cr}$ and $O_i$ 
both have formation energies of 2.0 eV, indicating that both
defects have near equal concentrations. For both defects, neutral charge 
state is the dominant charge state under \textit{p}-doped conditions.

\subsection{Interstitial Diffusion}
\subsubsection{Cr$_i$ diffusion}
\begin{wrapfigure}{r}{0.6\textwidth}
  \begin{subfigure}{0.3\textwidth}
    \centering
    \includegraphics[width=\linewidth]{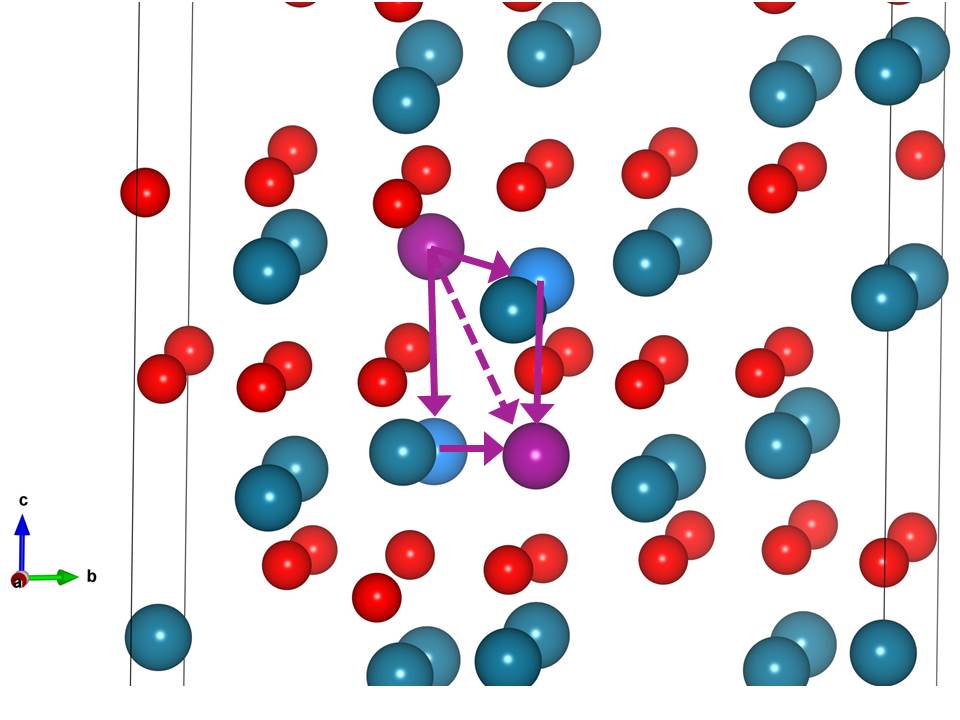}
    \caption{[221]}
    \label{fig:Crinter221diffpath}
  \end{subfigure}%
  \begin{subfigure}{0.3\textwidth}
    \centering
    \includegraphics[width=0.8\linewidth]{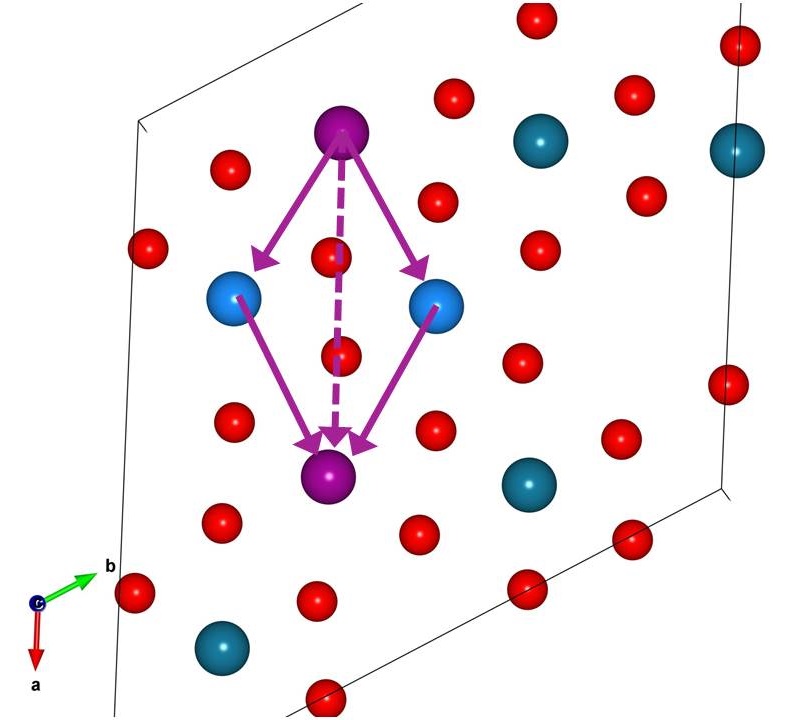}
    \caption{[100]}
    \label{fig:Crinter100diffpath}
  \end{subfigure}
  \caption{Migration pathways considered for Cr interstitial diffusion along (a) [221] (diagonal to corundum lattice) and (b) [100] directions. Magenta spheres represent the initial and final Cr interstitial positions.  The dotted line represent the direct hopping pathway, which could not be evaluated. The solid lines represent the paths, where the blue spheres hop to the final magenta sphere and create an intermediary defect complex comprising of interstitial and Frenkel defect. Teal and red spheres denote the regular Cr and O ions.  Plot (a) is oriented along side view and plot (b) is oriented along top view for clarity. }
\label{fig:Crinterdiffpathways}
\end{wrapfigure}
We have studied the diffusion of Cr and O interstitials in  charge states 
ranging from [0, 3] and [-2, 0], respectively.  The preferential diffusion 
pathways for both Cr and O interstitials are identified to be either in the 
basal plane  or along the diagonal of the standard corundum lattice.

Figure~\ref{fig:Crinterdiffpathways} illustrates the Cr diffusion 
pathways  studied. In this figure the magenta spheres  represent the 
initial and final   Cr interstitial sites and the blue spheres  
represent Cr ions affected by interstitial diffusion 
process.  Interstitial diffusion typically involves
two mechanisms: direct and interstitialcy diffusion. In the direct mechanism,
shown as dotted lines in Figure~\ref{fig:Crinterdiffpathways}, a Cr ion 
hops from one interstitial site to an adjacent interstitial site 
squeezing through the 
ions in the regular lattice. In the interstitialcy mode, a Cr interstitial  
knocks off a Cr ion in the regular Cr sublattice (represented by one of the two blue spheres) 
and occupies that sublattice site. The displaced
Cr ion goes to a neighboring unoccupied interstitial site. We investigated both
the pathways using CI-NEB method. However, either 
convergence could not be obtained in CI-NEB calculations, or the resulting 
barrier energies were very high (greater than 5 eV). This led us 
to investigate an alternative interstitial diffusion mechanism. 
In this mechanism,  diffusion is accomplished in two stages and 
involves an intermediary metastable  configuration. In the first 
stage, either one of the Cr ions along the diffusion path (represented by 
blue spheres)  jumps to the final interstitial 
position. This jump creates a defect complex comprising a Cr Frenkel pair and 
a Cr interstitial. This defect complex is quite similar to the intermediate 
triple defect (Cr Frenkel pair + Cr vacancy) involved in vacancy mediated 
Cr diffusion along c-axis~\cite{medasani2017}. During the second step,  Cr in 
the initial interstitial site jumps to the vacated Cr lattice site, 
thereby completing the 
diffusion process. This mechanism also resulted in lower barrier energies 
(for example, by {\raise.17ex\hbox{$\scriptstyle\sim$}}2 eV for 
$Cr_i^3$ along [221])
compared to direct and interstitialcy mechanisms in the few cases where 
CI-NEB calculations of direct and interstitialcy mechanisms 
converged implying 
that it is the true Cr interstitial diffusion mechanism in \ce{Cr2O3}. It 
may also represent one of the modes of cations interstitial diffusion 
in other corundum sesquioxides. For both the interstitialcy
and the proposed mechanisms, the atoms involved in the diffusion and 
their initial  and final positions are the same. The difference is 
that in the 
interstitialcy mechanism both atoms move concurrently, while
in the mechanism proposed in this work, they move consecutively.

\begin{table}
    \centering
    \caption{Migration barriers  for Cr interstitial mediated diffusion. $q$ denotes interstitial charge state.}
\begin{tabular}{lcccccccccccc}
    \hline
    Diffusion Path & \multicolumn{4}{c} {Length (\AA)} & \multicolumn{4}{c} {Attempt Frequency (THz)} & \multicolumn{4}{c} {Migration barrier (eV)} \\
    ($q=$)    &  0 & 1 & 2 & 3   &  0   &  1   &  2     &  3    &  0    & 1    & 2    & 3 \\
    \hline
    {[100]}   & 4.96 & 4.97 & 4.99 & 4.99 &  --  & --   & --     & --    &  4.04 & 5.53 & 8.21 & 5.2\\
    {[221]}   & 3.81 & 3.81 & 3.74 & 3.74 & 5.39 & 8.71 & 10.23  & 76.89 &  2.26 & 1.86 & 2.62 & 3.69 \\
    \hline
\end{tabular}
\label{tab:Crinterbarrieren}
\end{table}

\begin{wrapfigure}{r}{0.7\textwidth}
  \begin{subfigure}{\linewidth}
   \centering
    \includegraphics[width=0.5\linewidth]{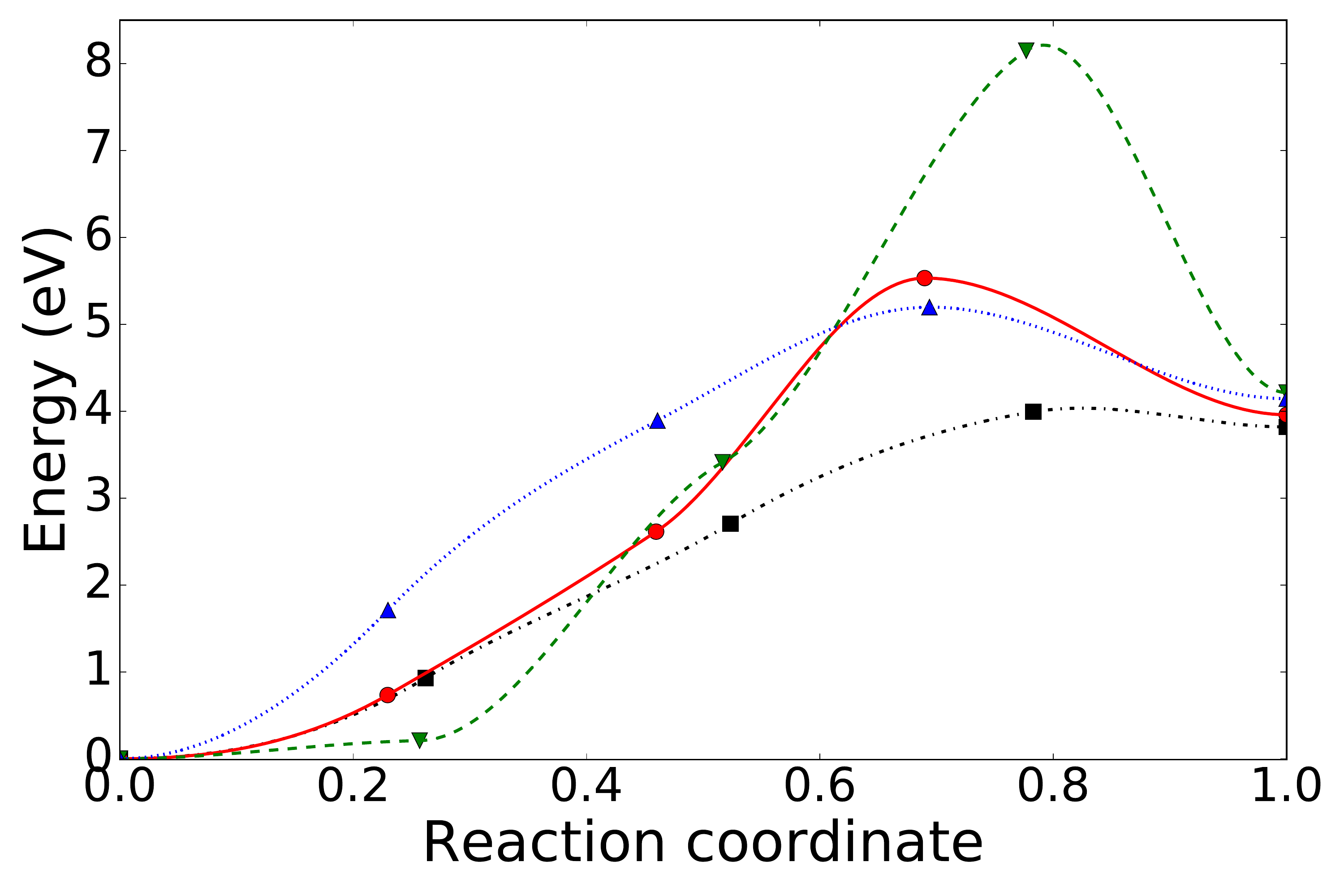}%
    \includegraphics[width=0.5\linewidth]{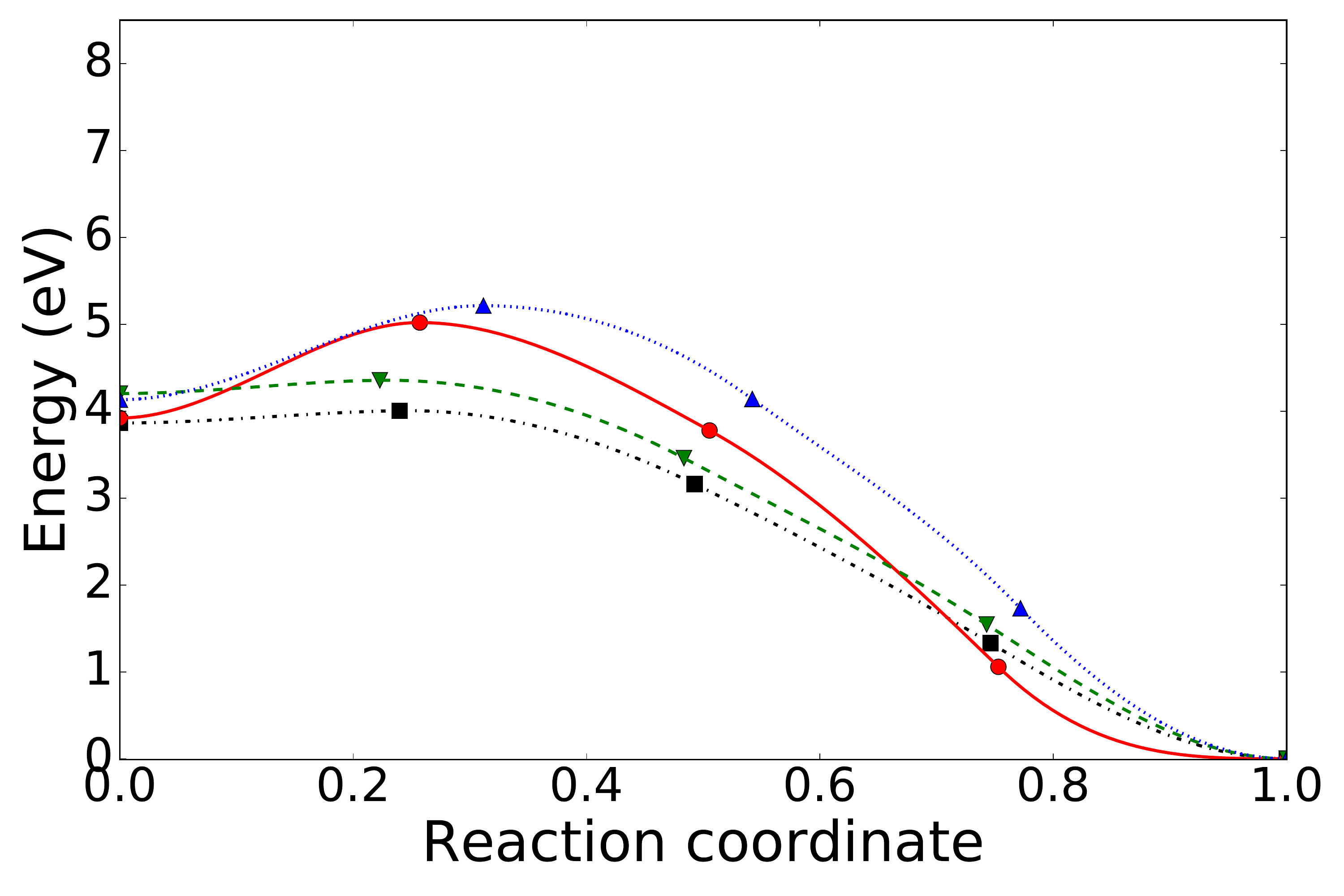}
    \caption{[100]}
    \label{fig:Crinterdiffbarrier100}
  \end{subfigure}
  \begin{subfigure}{\linewidth}
    \centering
    \includegraphics[width=0.5\linewidth]{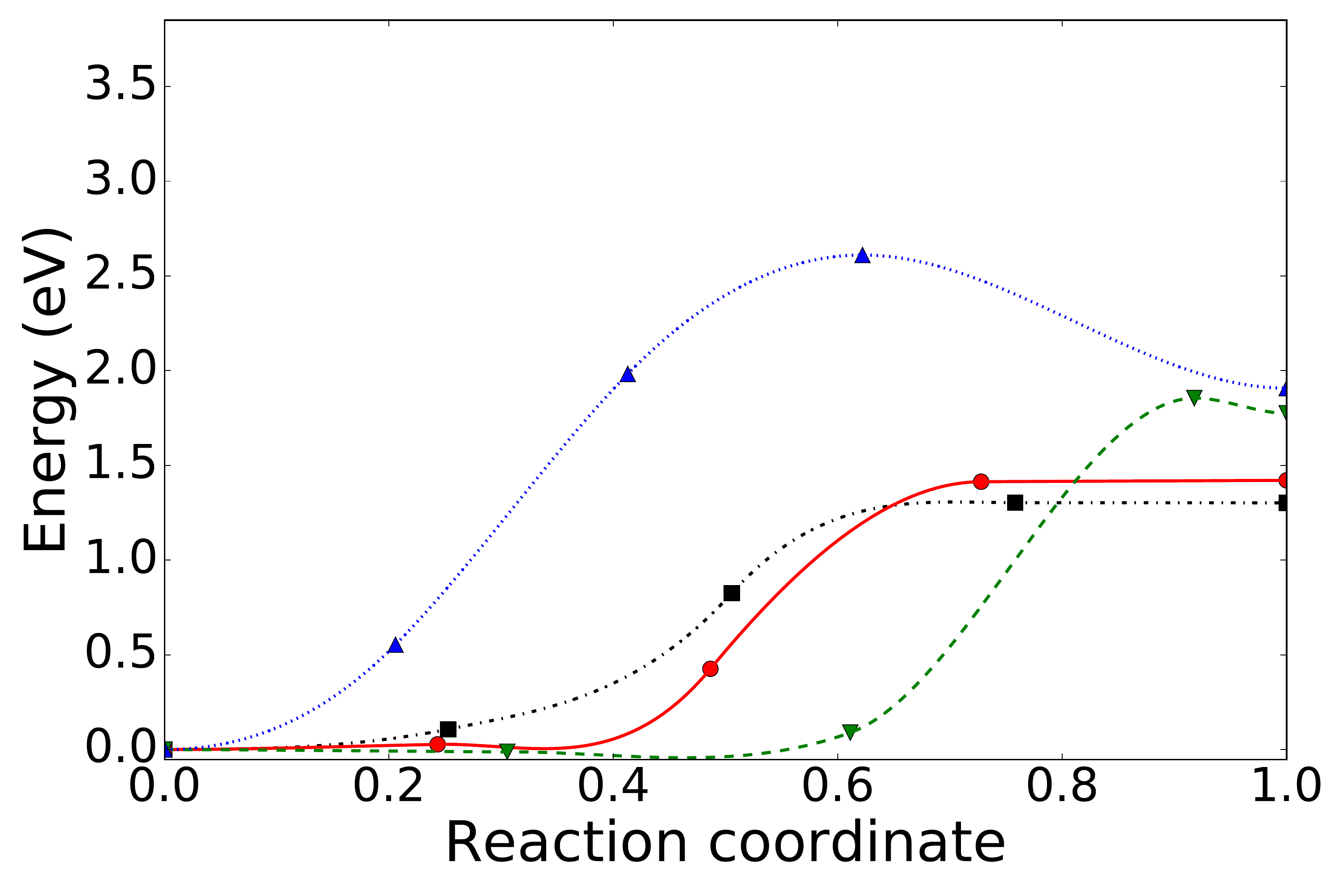}%
    \includegraphics[width=0.5\linewidth]{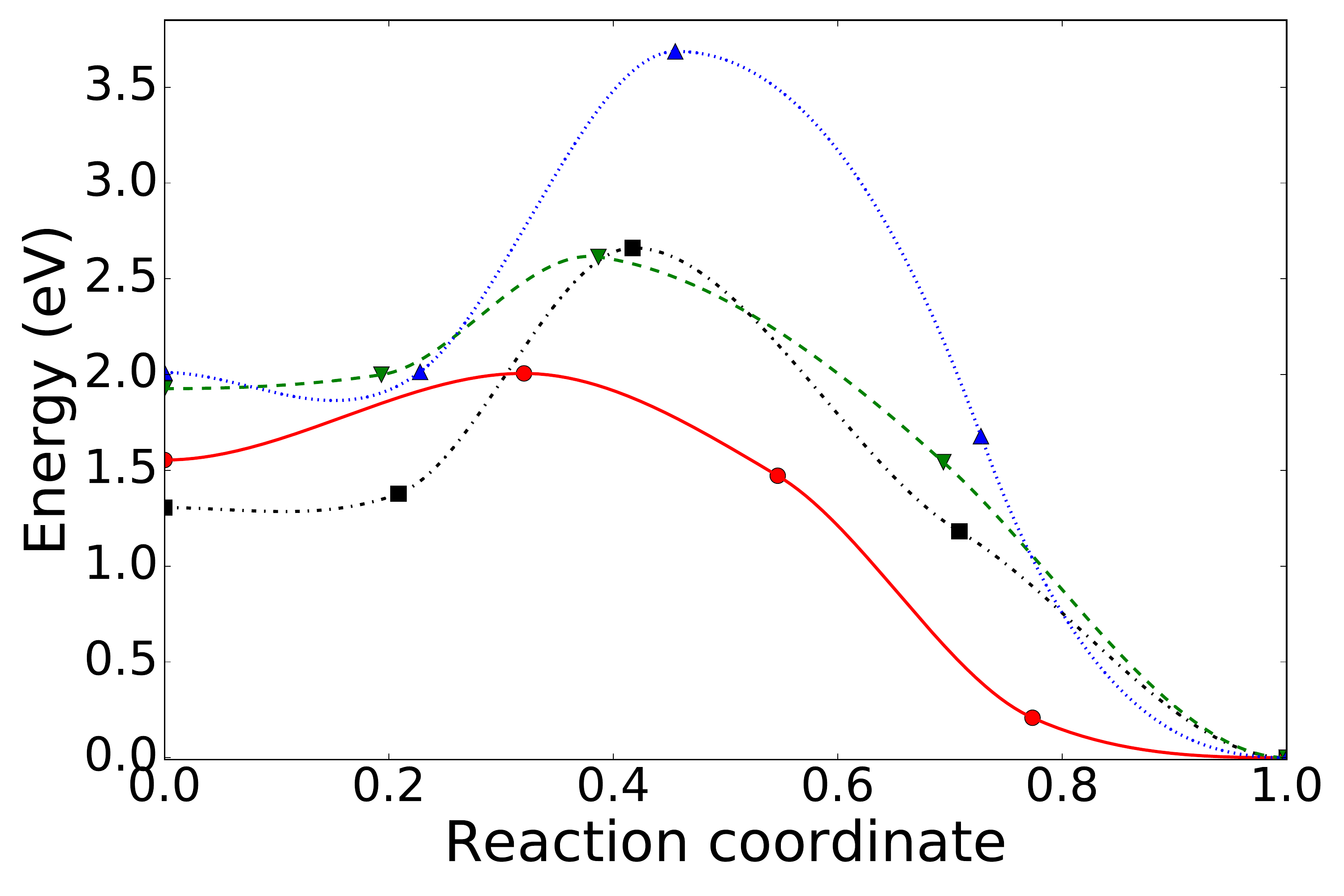}
    \caption{[221]}
    \label{fig:Crinterdiffbarrier221}
  \end{subfigure}
\caption{Migration barriers of Cr interstitial diffusion along (a) [100] and  (b) [221] pathways. Black, red, green, and blue lines represent ${Cr}_i^0$, $Cr_i^1$, $Cr_i^2$, and $Cr_i^3$, respectively.}  
\label{fig:Crinterdiffbarrier}
\end{wrapfigure}
Figure~\ref{fig:Crinter221diffpath} shows the diffusion pathway for Cr diffusion 
along the  [221]  direction. Diffusion involves either the Cr ion 
adjacent to the interstitial in the same Cr bilayer or the Cr ion in the 
next bilayer hopping to the final interstitial site. Depending on the Cr 
ion involved, elementary diffusion takes place either along the c-axis or 
along the basal plane, respectively. 
The subsequent jump by the second ion is then along the basal plane or 
the c-axis, respectively.  
Figure~\ref{fig:Crinter100diffpath} shows the diffusion pathway for Cr 
diffusion along basal plane (shown for [100] direction but also applies to [010]
direction). In the first stage, one of the six neighboring Cr ions of the 
occupied interstitial  jumps to one of the two available unoccupied interstitial 
sites. A total of 4 interstitial sites (two along [100] and two 
along [010]) are  available  for all the neighboring Cr ions. 

NEB barrier energy plots for Cr diffusion along
[100] and [221] directions are shown in
Figure~\ref{fig:Crinterdiffbarrier}. The left hand side 
and right hand side plots correspond to the first and second stages
of the Cr interstitial diffusion, respectively.
The plots reveal that Cr interstitials preferentially diffuse along the 
diagonal of the corundum lattice, in contrast to Cr vacancies 
that diffuse along the basal plane~\cite{medasani2017}.

The calculated barrier energies for Cr diffusion along [100] reveal that $Cr_i^0$ diffusion incurs
the lowest barrier of 4.04 eV  (Figure~\ref{fig:Crinterdiffbarrier100} and Table~\ref{tab:Crinterbarrieren}) 
followed by that for $Cr_i^3$, $Cr_i^1$, and $Cr_i^2$. Out of the interstitials studied, $Cr_i^1$, 
$Cr_i^2$ have higher barrier energies for the first stage 
of [100] diffusion compared to that for the second stage. In contrast, 
$Cr_i^0$ and $Cr_i^3$  exhibit nearly equal barrier 
energies for both stages  of diffusion.
Assuming the metastable defect complex as reference point, the nearly 
equal barrier energies imply that  either of the Cr interstitials in 
the initial and final positions can jump to the vacant Cr site with 
equal probability. However, jumping back of the Cr interstitial in the 
final position does not result in  net diffusion. Hence a prefactor 
of 0.5 needs to be added to the diffusion coefficient for $Cr_i^0$ 
and $Cr_i^3$.

Simulations of Cr interstitial diffusion along [221] suggest that
$Cr_i^1$ has the lowest diffusion barrier
of 1.86 eV (Figure~\ref{fig:Crinterdiffbarrier221} and 
Table~\ref{tab:Crinterbarrieren}) followed by $Cr_i^2$, 
$Cr_i^0$, and $Cr_i^3$. In contrast to Cr diffusion along [100] direction, [221] diffusion 
is characterized by relatively higher barrier energies for the second stage of the 
diffusion when compared to the first stage. However, we believe this difference is solely due to our choice of 
the ion (from the two ions available) involved in the first stage of the diffusion. 

Compared to Cr vacancies, which have the lowest diffusion barrier 
energies of 2.01 eV ($V_{Cr}^{-3}$ along basal plane)~\cite{medasani2017},
Cr interstitials 
have even lower barrier energies of 1.86 eV ($Cr_i^1$ along 
[221]).  The 0.15 eV lower barrier implies that Cr 
interstitials are more mobile than Cr vacancies by two 
orders of magnitude at room  temperature.
At very high temperatures (around 1200 K), the 
difference in the energy barriers becomes comparable 
to the  thermal energy and both Cr interstitials and vacancies have effectively equal mobilities.

It would be of interest to the community to identify physical factors that 
could be used either as descriptors to predict the barrier energies or to explain 
the variations in the barrier energies for different charge states.
Lei and Wang~\cite{Lei2015}, in their study of barrier energies of 
vacancies in \ce{Al2O3}, identified a correlation between barrier energies and the 
change in  defect levels of the vacancies between transition and ground 
states. Following that approach, we compared the barrier energies with the 
differences in the defect levels (Figure S6 of SI for transition 
state DOS) and also the differences in the electrostatic potentials 
(Table S6 of SI) experienced by the diffusing ion at the transition  
states of the [221] pathway and the interstitial ground states for $Cr_i$ 
diffusion. Unlike the case of \ce{Al2O3}, in \ce{Cr2O3} no such 
correlation could be found between the barrier energies of $Cr_i$ and 
either of the two physical  factors examined. Similarly, no such correlation 
was found for $O_i$ also, which are discussed in the following section (see Figures S7-S9 and Table S6 in SI).
The barrier energies for both $Cr_i$ and $O_i$ are a result of complex interplay of 
variations in local geometrical distortion, electrostatic potentials, and  
electronic correlation effects. 

\subsubsection{O$_i$ diffusion}
\begin{wrapfigure}{r}{0.5\textwidth}
    \vspace{-15pt}
  \begin{subfigure}{0.5\linewidth}
    \centering
    \includegraphics[width=\linewidth]{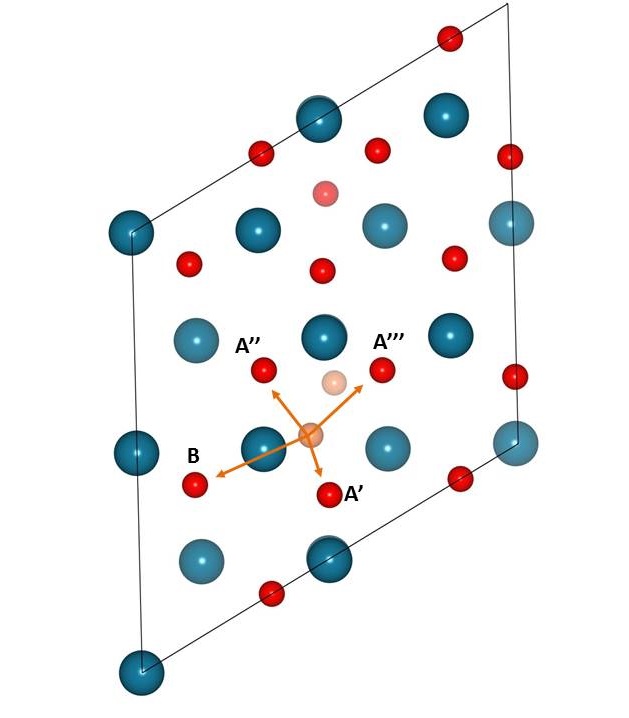}
    \caption{[221]}
    \label{fig:Ointer221diffpath}
  \end{subfigure}%
  \begin{subfigure}{0.5\linewidth}
    \centering
    \includegraphics[width=\linewidth]{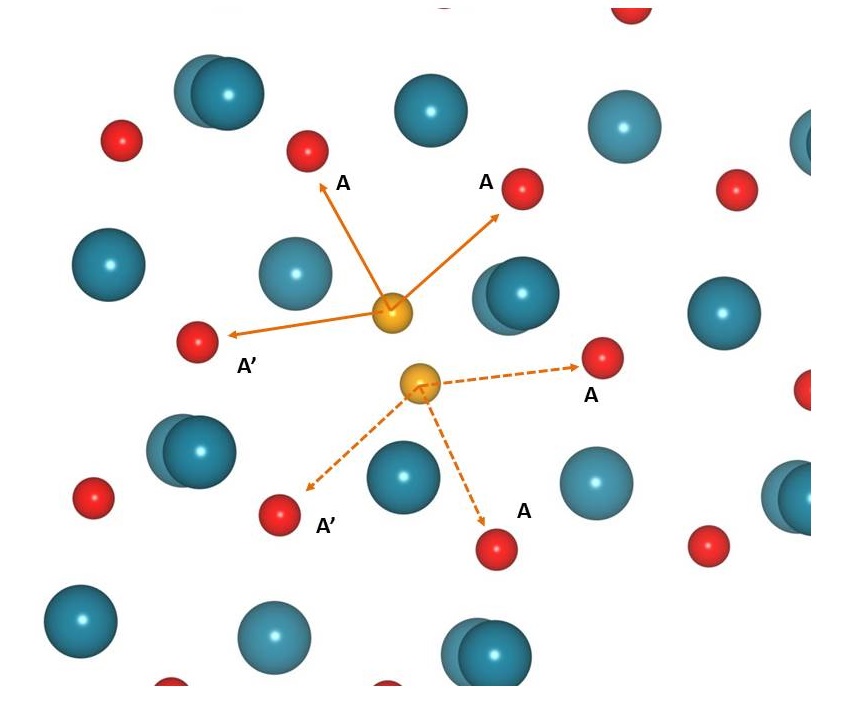}
    \caption{[100]}
    \label{fig:Ointer100diffpath}
  \end{subfigure}
    \vspace{-10pt}
  \caption{Migration pathways identified for O interstitial diffusion (via 
  bond order switching) along (a) [221] (diagonal to corundum lattice) and 
  (b) [100] directions. Orange spheres represent the O interstitial in 
  dumbbell configuration. The view is from top (along \textit{c}-axis)
  for both the plots. Teal  spheres denote the Cr  ions. For plot (a),  dark red spheres represent O ions that
  are in the upper adjacent O-layer neighboring the O-layer 
  consisting the  O interstitial ions. For reference, one O-ion in the same layer of the interstitial O ions is shown in lighter red.  In plot 
  (b), the red spheres represent the O ions in the O-layer consisting of 
  the O interstitial.  }
\label{fig:Ointerdiffpathways}
    \vspace{-15pt}
\end{wrapfigure}
An O interstitial forms a dumbbell configuration oriented along [221] 
around a regular O  site. O ion diffusion is accomplished by bond 
switching, where one of the interstitial
O ions jumps to one of the nearest O sites and forms a dumbbell 
configuration  with the lattice O ion. The remaining 
O ion  at the initial  interstitial location returns to its regular 
lattice position. 

\begin{table}
    \centering
    \caption{Migration barriers for O interstitial mediated diffusion. $q$ denotes vacancy charge state.}
\begin{tabular}{lccccccccc}
    \hline
    Diffusion Path & \multicolumn{3}{c} {Length (\AA)} & \multicolumn{3}{c} {Attempt Frequency (THz)} &  \multicolumn{3}{c} {Migration barrier (eV)}  \\
    ($q=$)  &  0 & -1 & -2 &  0 & -1 & -2     &  0 & -1 & -2  \\
    \hline
    {[100]}       & 2.12 & 2.53 & 2.09 & 54.98 & 2.73 & 6.35 & 1.52 & 2.12 & 0.30 \\
    {[221]} short & 2.34 & 2.25 & 2.23 & 27.15 & 2.13 & 6.77 & 1.27 & 0.21 & 0.29 \\
    {[221]} long  & 3.45 & 3.2  & 3.12 &  --   & 6.34 &48.12 &  --  & 1.63 & 0.58 \\
    \hline
\end{tabular}
\label{tab:Ointerbarrieren}
\end{table}

O diffusion pathways considered in this study are shown in
Figure~\ref{fig:Ointerdiffpathways}. The red spheres in
Figure~\ref{fig:Ointer221diffpath} represent the O ions in the 
adjacent to the defect plane O layer and the labeled ions are the nearest  to the upper O 
interstitial ion (dark orange sphere).  The ions labeled 
with A$^\prime$, A$^{\prime\prime}$, and A$^{\prime\prime\prime}$ represent
three closest ions equidistant to the center of the dumbbell configuration in a perfect 
crystal. Of these A$^\prime$ is the closest site, followed by A$^{\prime\prime}$ and A$^{\prime\prime\prime}$.
The sphere labeled as B represents the next closest  O ion at around 
3.1 -- 3.5~\AA{} from the dumbbell center, depending on the interstitial charge  state.  
We computed the barrier energies associated with $O_i$ diffusion to
A$^\prime$ and B sites, denoted as [221]-short and [221]-long, 
respectively (Table.~\ref{tab:Ointerbarrieren}).
In  Figure~\ref{fig:Ointer100diffpath}, the red spheres 
represent O ions in the same layer as the interstitial 
ions. Out of six nearest ions the ions represented by A are 
equidistant from the O interstitial and those denoted 
by A$^\prime$ are slightly farther away. The barrier energies were computed for O 
diffusion from interstitial site to A site.

\begin{wrapfigure}{r}{0.4\textwidth}
  \begin{subfigure}{\linewidth}
    \centering
    \includegraphics[width=\linewidth]{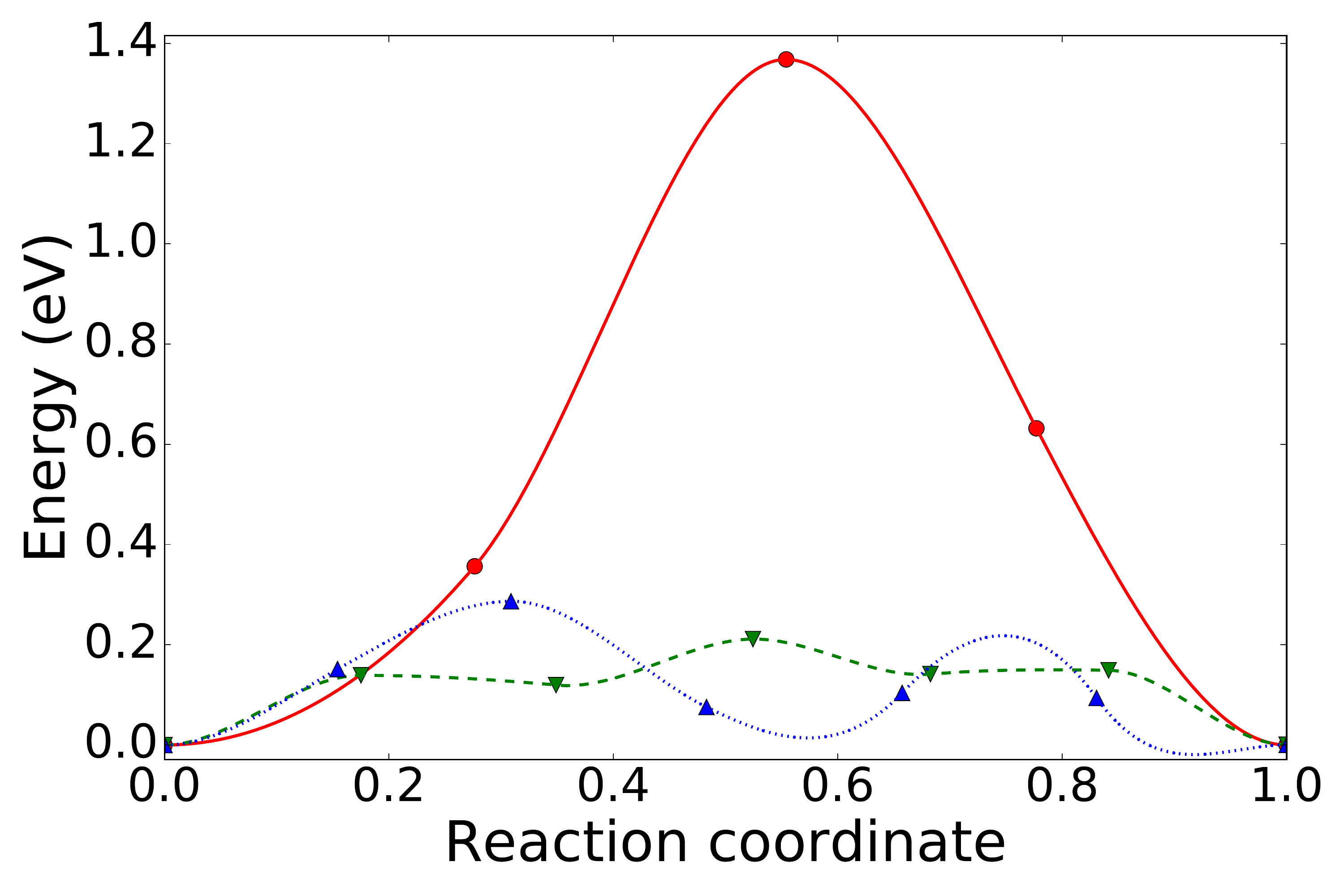}
    \caption{[221] short}
    \label{fig:Odiffpath221short}
  \end{subfigure}
  \begin{subfigure}{\linewidth}
    \centering
    \includegraphics[width=\linewidth]{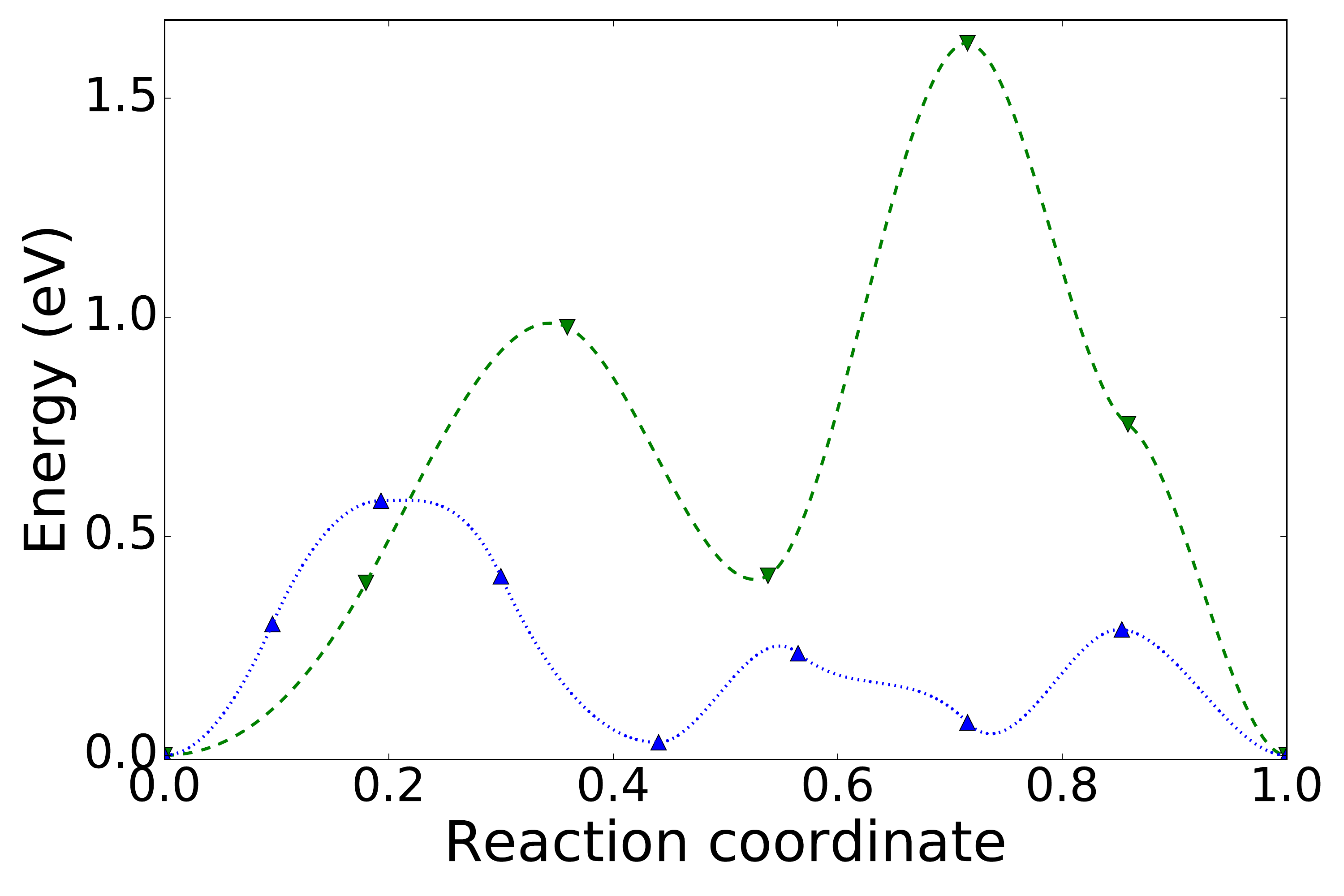}
    \caption{[221] long}
    \label{fig:Odiffpath221long}
  \end{subfigure}
  \begin{subfigure}{\linewidth}
    \centering
    \includegraphics[width=\linewidth]{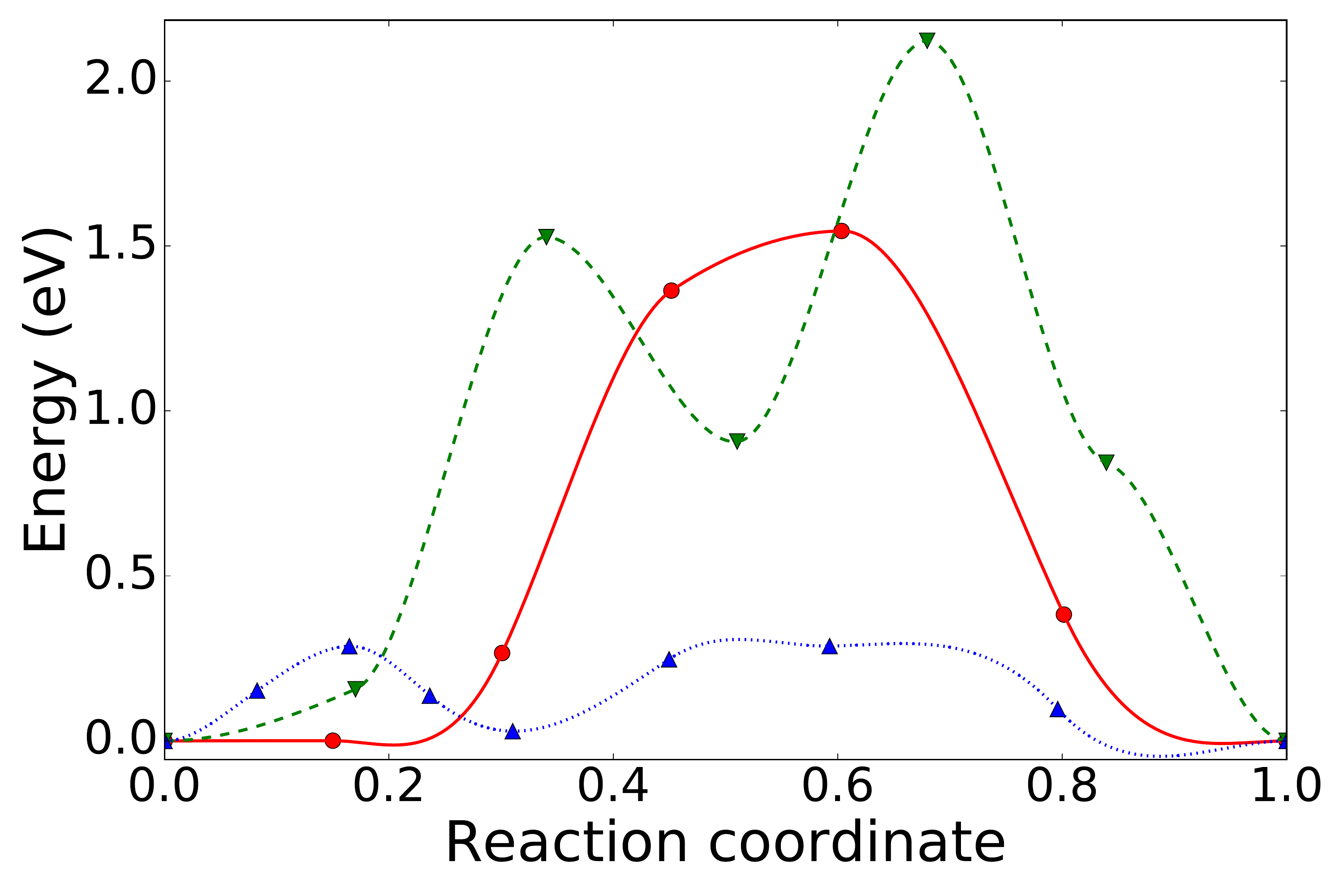}
    \caption{[100]}
    \label{fig:Odiffpath100}
  \end{subfigure}
  \caption{Migration barriers for O diffusion along (a) [221] short, (b) [221] long, and (c) [100] pathways. Red, green, and blue lines represent $O_i^0$, $O_i^{-1}$, and $O_i^{-2}$ respectively. }
\label{fig:Odiff}
\vspace{-10pt}
\end{wrapfigure}
 O diffusion calculations show that O migration energies 
 are highly dependent on the interstitial charge
and the diffusion pathway (Figure~\ref{fig:Odiff}).
$O_i^{-2}$ has the lowest barrier energies ranging 
from 0.29 to 0.58 eV. Further $O_i^{-2}$ diffusion pathway encompasses
one metastable state
along [221]-short and [100] pathways and two metastable states in the 
[221]-long pathway. 
The metastable state corresponds to the diffusing O interstitial ion 
breaking bond with the other O-ion in the dumbbell and occupying
the 6-O coordinated octahedral site. 
The identical phonon DOS plots of 
the transition states (Fig. S1 in SI)  and the nearly equal barrier energies
for [221]-short and [100] pathways indicate that both pathways 
have a common transition state for -2 charge state. 
$O_i^{-1}$ has a significantly lower barrier energy 
of 0.21 eV along the [221]-short pathway when compared to the 1.63 eV 
and 2.12 eV energies along [221]-long and [100] pathways respectively. 
Neutral $O_i$ on the other hand exhibits comparatively higher barrier 
energies around 1.5 eV for [221]-short and [100] pathways. We could not obtain
convergence for $O_i^0$ diffusion along [221]-long path.

When compared to the O vacancies, O interstitials exhibit significantly lower
barrier energies for diffusion. The lowest barrier energies of O vacancies are of the 
order of 1.2 -- 1.5 eV for $V_O^2$~\cite{medasani2017}, whereas $O_i^{-1}$ and 
$O_i^{-2}$  have the lowest barrier energies of the order of 0.2 and 0.3 eV 
respectively. For the undoped \ce{Cr2O3}, however, consideration
has to be given to neutral vacancies and neutral oxygen interstitials, because of their
higher concentration at $E_F = E_g/2$. The neutral O interstitials are 
also significantly more mobile than neutral O vacancies with a 0.8 eV difference in 
the lowest barrier energies. It is noteworthy that under irradiation conditions, 
characterized by considerable and nearly equal in concentrations of interstitials
and vacancies due to the 
formation of Frenkel defects, our results for barrier energies indicate that
O interstitials could be responsible for  oxygen transport in \ce{Cr2O3}.

\section{Summary\label{sec:conclusions}}
To develop a comprehensive model of defect-mediated diffusion processes in 
\ce{Cr2O3}, we computed the electronic, thermodynamic, and diffusion 
properties of interstitials in \ce{Cr2O3}. This work is an extension 
to our previous study of vacancies in \ce{Cr2O3}.  
The formation energies of Cr and  O interstitials in various charge states 
were evaluated for Cr-rich and O-rich conditions using the thermodynamic  
formalism proposed by Zhang and Northrup~\cite{zhang1991}.
Our results indicate that Cr and O interstitials are deep defects, and 
contrary to vacancies, they 
do not generate any considerable local  structural distortions. 

O interstitials form dumbbell configurations and O diffusion is accomplished
via bond switching mechanism.
The computed migration barrier energies reveal that similar to vacancies, 
O interstitials are more mobile than Cr interstitials. Among different charge 
states of O interstitials, $O_i^{-2}$ has the highest mobility for nearly all pathways. 
Overall, O interstitials
are more mobile, i.e. incur lower barrier energies, than O vacancies.

Cr interstitials occupy the empty octahedral sites in the Cr sub-lattice
and our calculations indicate that the preferred path  for Cr interstitial diffusion 
is along [221]  direction. The diffusion is accomplished via an 
intermediate metastable interstitial  triple-defect configuration 
comprising a Cr interstitial and a Cr Frenkel defect in the adjacent Cr bilayer.
Cr diffusion  from one interstitial site to another site comprises two stages, 
 with both basal and c-axis oriented diffusion happening in individual stages.
 The mobility of Cr interstitials is slightly higher but comparable to that of Cr vacancies.

The  diffusion mechanisms of self-interstitials in \ce{Cr2O3} revealed in the 
present study  can be used to 
understand the efficacy of \ce{Cr2O3} as a passivation layer and the corrosion
mechanisms in Cr alloys under various physical conditions including irradiated conditions.

\section*{Acknowledgement}

This work was supported by the U.S. Department of Energy, Office of Science, Basic Energy Sciences, Materials Sciences and Engineering Division. Simulations were performed using PNNL Institutional Computing facility. PNNL is a multiprogram National Laboratory operated by Battelle for the U.S. Department of Energy under Contract DE-AC06-76RLO 1830.





\begin{thebibliography}{10}
  \def\url#1{\texttt{#1}}

\bibitem{naceimpact}
G.~Koch, J.~Varney, N.~Thompson, O.~Moghissi, M.~Gould, J.~Payer,
  {Nace
  international measures of prevention, application, and economics of corrosion
  technologies study}, [Online; accessed 28-Feb-2018] (2016).
\newline\url{http://impact.nace.org/documents/Nace-International-Report.pdf}

\bibitem{infacon1992}
H.~W. Glea (Ed.), Vol.~2 of Proceedings of the 1st International Chromium Steel
  and Alloys Congress, SAIMM, 1992.

\bibitem{rooyen1992}
G.~T. Van-Rooyen, The potential of chromium as an alloying element, in: H.~W.
  Glea (Ed.), Chromium Steel and Alloys, Vol.~2 of Proceedings of the 1st
  International Chromium Steel and Alloys Congress, SAIMM, Cape Town, South
  Africa, 1992, pp. 43--47.

\bibitem{asm}
   S.~D. Cramer, J.~B.~S.~Covino (Eds.), Corrosion: Materials, Vol. 13B, and Corrosion: Environment and
  Industries, Vol. 13C of ASM Handbook; ASM International: Materials Park, OH, USA, 2005. 

\bibitem{was2005}
G.~S. Was, J.~T. Busby,
  {Role
  of irradiated microstructure and microchemistry in irradiation-assisted
  stress corrosion cracking}, Philos. Mag. 85~(4-7) (2005) 443--465.
\newblock doi:10.1080/02678370412331320224.

\bibitem{zinkle2009}
S.~J. Zinkle, J.~T. Busby,
  {Structural
  materials for fission and fusion energy}, Mater. Today 12~(11) (2009) 12 --
  19.
\newblock 
  doi:10.1016/S1369-7021(09)70294-9.

\bibitem{schreiber2013}
D.~Schreiber, M.~Olszta, D.~Saxey, K.~Kruska, K.~Moore, S.~Lozano-Perez,
  S.~Bruemmer,
  {Examinations
  of oxidation and sulfidation of grain boundaries in alloy 600 exposed to
  simulated pressurized water reactor primary water}, Microscopy and
  Microanalysis 19 (2013) 676--687.
\newblock 
  doi:10.1017/S1431927613000421.

\bibitem{schreiber2014}
D.~Schreiber, M.~Olszta, S.~Bruemmer,
  {Grain
  boundary depletion and migration during selective oxidation of cr in a ni-5cr
  binary alloy exposed to high-temperature hydrogenated water}, Scripta
  Materialia 89 (2014) 41 -- 44.
\newblock 
  doi:10.1016/j.scriptamat.2014.06.022.

\bibitem{kofstad1982}
P.~Kofstad, K.~Lillerud, Chromium transport through \ce{Cr2O3} scales i. on
  lattice diffusion of chromium, Oxidation of metals 17~(3-4) (1982) 177--194.

\bibitem{sabioni1992a}
A.~Sabioni, B.~Lesage, A.~Huntz, J.~Pivin, C.~Monty, {Self-diffusion in
  \ce{Cr2O3} I. Chromium diffusion in single crystals}, Philosophical Magazine
  A 66~(3) (1992) 333--350.

\bibitem{sabioni1992b}
A.~Sabioni, A.~Huntz, F.~Millot, C.~Monty, {Self-diffusion in \ce{Cr2O3} II.
  Oxygen diffusion in single crystals}, Philosophical Magazine A 66~(3) (1992)
  351--360.

\bibitem{sabioni1992c}
A.~Sabioni, A.~Huntz, F.~Millot, C.~Monty, {Self-diffusion in \ce{Cr2O3} III.
  Chromium and oxygen grain-boundary diffusion in polycrystals}, Philosophical
  Magazine A 66~(3) (1992) 361--374.

\bibitem{hoshino1983}
K.~Hoshino, N.~Peterson, Cation self-diffusion in \ce{Cr2O3}, Journal of the
  American Ceramic Society 66~(11).

\bibitem{latanision1967}
R.~Latanision, R.~Staehle, Stress corrosion cracking of iron--nickel--chromium
  alloys, Tech. rep., Ohio State Univ. Research Foundation, Columbus (1967).

\bibitem{Schmucker2016}
E.~Schmucker, C.~Petitjean, L.~Martinelli, P.-J. Panteix, S.~B. Lagha,
  M.~Vilasi,
  {Oxidation
  of ni-cr alloy at intermediate oxygen pressures. i. diffusion mechanisms
  through the oxide layer}, Corrosion Science 111 (2016) 474--485.
\newblock 
  doi:10.1016/j.corsci.2016.05.025.

\bibitem{cao2017}
P.~Cao, D.~Wells, M.~P. Short, Anisotropic ion diffusion in
  $\alpha$-\ce{Cr2O3}: an atomistic simulation study, Phys. Chem. Chem. Phys.
  19 (2017) 13658.

\bibitem{Lebreau2014}
F.~Lebreau, M.~M. Islam, B.~Diawara, P.~Marcus,
  {Structural, magnetic, electronic,
  defect, and diffusion properties of \ce{Cr2O3}: A dft+u study}, J. Phys.
  Chem. C 118~(31) (2014) 18133--18145.
\newblock doi:10.1021/jp5039943.

\bibitem{medasani2017}
B.~Medasani, M.~L. Sushko, K.~M. Rosso, D.~K. Schreiber, S.~M. Bruemmer,
  {Vacancies and
  vacancy-mediated self diffusion in cr2o3: A first-principles study}, J. Phys.
  Chem. C 121~(3) (2017) 1817--1831.
\newblock 
  doi:10.1021/acs.jpcc.7b00071.

\bibitem{gray2016}
C.~Gray, Y.~Lei, G.~Wang, {Charged
  vacancy diffusion in chromium oxide crystal: Dft and dft+u predictions}, J.
  Appl. Phys. 120~(21) (2016) 215101.
\newblock 
  doi:10.1063/1.4970882.

\bibitem{Vaari2015}
J.~Vaari,
  {Molecular
  dynamics simulations of vacancy diffusion in chromium(iii) oxide, hematite,
  magnetite and chromite}, Solid State Ion. 270 (2015) 10 -- 17.
\newblock
  doi:10.1016/j.ssi.2014.11.027.

\bibitem{laturomain2017}
L.~Latu-Romain, Y.~Parsa, S.~Mathieu, M.~Vilasi, A.~Galerie, Y.~Wouters,
  {Towards
  the growth of stoichiometric chromia on pure chromium by the control of
  temperature and oxygen partial pressure}, Corrosion Science 126 (2017) 238 --
  246.
\newblock 
  doi:10.1016/j.corsci.2017.07.005.

\bibitem{hohenberg}
P.~Hohenberg, W.~Kohn,
  {Inhomogeneous
  electron gas}, Phys. Rev. 136 (1964) B864--B871.
\newblock 
  doi:10.1103/PhysRev.136.B864.

\bibitem{kohn}
W.~Kohn, L.~J. Sham,
  {Self-consistent
  equations including exchange and correlation effects}, Phys. Rev. 140 (1965)
  A1133--A1138.
\newblock 
  doi:10.1103/PhysRev.140.A1133.

\bibitem{dftprimer}
J.~P. Perdew, S.~Kurth, in: C.~Fiolhais, F.~Nogueira, M.~Marques (Eds.), A
  Primer in Density Functional Theory, Springer, Berlin, 1990.

\bibitem{freysoldt2014}
C.~Freysoldt, B.~Grabowski, T.~Hickel, J.~Neugebauer, G.~Kresse, A.~Janotti,
  C.~G. Van~de Walle,
  {First-principles
  calculations for point defects in solids}, Rev. Mod. Phys. 86 (2014)
  253--305.
\newblock 
  doi:10.1103/RevModPhys.86.253.

\bibitem{pycdt}
D.~Broberg, B.~Medasani, N.~E. Zimmermann, G.~Yu, A.~Canning, M.~Haranczyk,
  M.~Asta, G.~Hautier,
  {Pycdt:
  A python toolkit for modeling point defects in semiconductors and
  insulators}, Computer Physics Communications 226 (2018) 165 -- 179.
\newblock 
  doi:10.1016/j.cpc.2018.01.004.

\bibitem{ong2012b}
S.~P. Ong, W.~D. Richards, A.~Jain, G.~Hautier, M.~Kocher, S.~Cholia,
  D.~Gunter, V.~L. Chevrier, K.~A. Persson, G.~Ceder, Python materials genomics
  (pymatgen) a robust, open-source python library for materials analysis,
  Comput. Mater. Sci. 68 (2013) 314.

\bibitem{zimmermann2016}
N.~E.~R. Zimmermann, M.~K. Horton, A.~Jain, M.~Haranczyk,
  {Assessing
  local structure motifs using order parameters for motif recognition,
  interstitial identification, and diffusion path characterization}, Frontiers
  in Materials 4 (2017) 34.
\newblock 
  doi:10.3389/fmats.2017.00034.

\bibitem{kresse93}
G.~Kresse, J.~Hafner,
  {\textit{Ab initio}
  molecular dynamics for liquid metals}, Phys. Rev. B 47 (1993) 558--561.
\newblock 
  doi:10.1103/PhysRevB.47.558.

\bibitem{kresse94}
G.~Kresse, J.~Hafner,
  {\textit{Ab initio}
  molecular-dynamics simulation of the
  liquid-metal\char21{}amorphous-semiconductor transition in germanium}, Phys.
  Rev. B 49 (1994) 14251--14269.
\newblock 
  doi:10.1103/PhysRevB.49.14251.

\bibitem{kresse96}
G.~Kresse, J.~Furthm\"uller,
  {Efficient iterative
  schemes for \textit{ab initio} total-energy calculations using a plane-wave
  basis set}, Phys. Rev. B 54 (1996) 11169--11186.
\newblock 
  doi:10.1103/PhysRevB.54.11169.

\bibitem{dftuliech}
A.~I. Liechtenstein, V.~I. Anisimov, J.~Zaanen,
  {Density-functional
  theory and strong interactions: Orbital ordering in mott-hubbard insulators},
  Phys. Rev. B 52 (1995) R5467--R5470.
\newblock 
  doi:10.1103/PhysRevB.52.R5467.

\bibitem{Rohrbach2004}
A.~Rohrbach, J.~Hafner, G.~Kresse,
  {\textbf{\textit{ab
  initio}} study of the (0001) surfaces of hematite and chromia: Influence of
  strong electronic correlations}, Phys. Rev. B 70 (2004) 125426.
\newblock 
  doi:10.1103/PhysRevB.70.125426.

\bibitem{zhang1991}
S.~B. Zhang, J.~E. Northrup, Chemical potential dependence of defect formation
  energies in {GaAs}: Application to ga self-diffusion, Phys. Rev. Lett. 67
  (1991) 2339--2342.

\bibitem{cineb}
G.~Henkelman, B.~P. Uberuaga, H.~J\'{o}nsson, A climbing image nudged elastic
  band method for finding saddle points and minimum energy paths, J. Chem.
  Phys. 113 (2000) 9901--9904.

\bibitem{phonopy}
A.~Togo, I.~Tanaka, First principles phonon calculations in materials science,
  Scr. Mater. 108 (2015) 1--5.

\bibitem{vesta}
K.~Momma, F.~Izumi, {{\it VESTA3} for three-dimensional visualization of
  crystal, volumetric and morphology data}, J. Appl. Crystallogr. 44 (2011)
  1272--1276.

\bibitem{huang2014}
B.~Huang, R.~Gillen, J.~Robertson,
  {Study of ceo2 and its native defects
  by density functional theory with repulsive potential}, The Journal of
  Physical Chemistry C 118~(42) (2014) 24248--24256.
\newblock 
  doi:10.1021/jp506625h.

\bibitem{kehoe2016}
A.~B. Kehoe, E.~Arca, D.~O. Scanlon, I.~V. Shvets, G.~W. Watson,
  {Assessing the
  potential of mg-doped cr 2 o 3 as a novel p -type transparent conducting
  oxide}, Journal of Physics: Condensed Matter 28~(12) (2016) 125501.

\bibitem{Carey2016}
J.~J. Carey, M.~Legesse, M.~Nolan,
  {Low valence cation doping of
  bulk cr2o3: Charge compensation and oxygen vacancy formation}, The Journal of
  Physical Chemistry C 120~(34) (2016) 19160--19174.
\newblock 
  doi:10.1021/acs.jpcc.6b05575.

\bibitem{lany2015}
S.~Lany, {Semiconducting
  transition metal oxides}, Journal of Physics: Condensed Matter 27~(28) (2015)
  283203.

\bibitem{freysoldt2009}
C.~Freysoldt, J.~Neugebauer, C.~G. Van~de Walle,
  {Fully ab
  initio finite-size corrections for charged-defect supercell calculations},
  Phys. Rev. Lett. 102 (2009) 016402.
\newblock 
  doi:10.1103/PhysRevLett.102.016402.

\bibitem{kumagai2014}
Y.~Kumagai, F.~Oba,
  {Electrostatics-based
  finite-size corrections for first-principles point defect calculations},
  Phys. Rev. B 89 (2014) 195205.
\newblock 
  doi:10.1103/PhysRevB.89.195205.

\bibitem{janotti2007}
A.~Janotti, C.~G. Van~de Walle, Native point defects in zno, Phys. Rev. B 76
  (2007) 165202.

\bibitem{Lei2015}
Y.~Lei, G.~Wang,
  {Linking
  diffusion kinetics to defect electronic structure in metal oxides:
  Charge-dependent vacancy diffusion in alumina}, Scripta Materialia 101 (2015)
  20 -- 23.
\newblock 
  doi:10.1016/j.scriptamat.2015.01.008.

\end{thebibliography}
\bibliographystyle{elsarticle-num}

\end{document}